\documentclass{aa}

\usepackage{graphicx,dblfloatfix}
\usepackage{txfonts}
\usepackage{hyperref}
\usepackage{color}

\begin{document}

   \title{Evolution and wave-like properties \\ of the average solar supergranule}

   \author{J. Langfellner \inst{1}
             \and
             A.~C. Birch \inst{1}
             \and
             L. Gizon \inst{1,2,3}}

   \institute{Max-Planck-Institut f\"ur Sonnensystemforschung,
             Justus-von-Liebig-Weg 3, 37077 G\"ottingen, Germany
             \and
             Georg-August-Universit\"at, Institut f\"ur Astrophysik,
               Friedrich-Hund-Platz 1, 37077 G\"ottingen, Germany
             \and
             Center for Space Science, NYUAD Institute, New York University Abu
             Dhabi, PO Box 129188, Abu Dhabi, UAE
             }

   \date{Received <date> / Accepted <date>}

% \abstract{}{}{}{}{} 
% 5 {} token are mandatory
 
  \abstract
  % context heading (optional)
   {Solar supergranulation presents us with many mysteries. For example, previous studies in spectral space found that supergranulation has wave-like properties.}
  % aims heading (mandatory)
   {Here we study, in real space, the wave-like evolution of the average supergranule over a range of spatial scales (from 10 to 80~Mm). We complement this by characterizing the evolution of the associated network magnetic field.}
  % methods heading (mandatory)
   {We use one year of data from the Helioseismic and Magnetic Imager (HMI) onboard the Solar Dynamics Observatory to measure horizontal near-surface flows near the solar equator 
   by applying time-distance helioseismology (TD) on Dopplergrams and granulation tracking (LCT) on intensity images. The average supergranule outflow (or inflow) is constructed by averaging over 10,000 individual outflows (or inflows). The contemporaneous evolution of the  magnetic field is studied with HMI line-of-sight observations.}
  % results heading (mandatory)
   {We confirm and extend previous measurements of the supergranular wave dispersion relation to angular wavenumbers in the range $50 < kR_\odot < 270$. We find a plateau for $kR_\odot > 120$.
   In real space, larger supergranules undergo oscillations with longer periods and lifetimes than smaller cells.
   We find excellent agreement between TD and LCT and obtain wave properties that are independent of the tracking rate. The observed network magnetic field follows the oscillations of the supergranular flows with a six-hour time lag. This behavior can be explained by computing the motions of corks carried by the supergranular flows.}
  % conclusions heading (optional)
   {Signatures of supergranular waves in surface horizontal flows near the solar equator can be observed in real space. These oscillatory flows control the evolution of the network magnetic field, in particular they explain the recently discovered east-west anisotropy of the magnetic field around the average supergranule. Background flow measurements that we obtain from Doppler frequency shifts do not favor shallow models of supergranulation.}
%}


   \maketitle
%

%%%%%%%%%%%%%%%%%%%%%%%%%%%%%%%%%%%%%%%%%%%%%%%%%%%%%%%%%%%%%%%%%%%%%%%%%%%%%%%%%

\section{Introduction}

The Sun exhibits convective motions at its surface on a range of scales, reaching from granulation to giant cells. Supergranulation \citep{rieutord_2010} is a pattern on intermediate scales (${\sim}30$~Mm) with horizontal velocities of a few hundred meters per second.
Even half a century after the discovery of supergranulation \citep{hart_1954}, its origin is still not understood, although many attempts have been made \citep[see, e.g.,][for an extensive list]{rieutord_2010}. Recently, \citet{cossette_2016} suggested, based on results from anelastic simulations, that supergranulation might be the largest spatial scale of convection that is driven by buoyancy. \citet{featherstone_2016}, on the other hand, argued that solar rotation might suppress convection on scales larger than supergranules. Other scenarios involve a prominent role of the magnetic field. For example, \citet{crouch_2007} and \citet{thibault_2012} consider that the network field could arise from random clustering of magnetic elements, without the need for supergranular flows. \citeauthor{crouch_2007} then hypothesize that the network might drive the supergranular flows.

Another prominent feature of the supergranular pattern is its super-rotation, i.e.~the pattern rotation rate is higher than that of granules and magnetic elements \citep[e.g.,][]{duvall_1980,snodgrass_1990,meunier_2007}. Using time-distance helioseismology and direct Doppler imaging, respectively, \citet{gizon_2003} and \citet{schou_2003} found signatures in the supergranular power spectrum that are characteristic of waves with periods of 6$-9$~days. At the equator, these waves were found to travel predominantly in the east-west direction, with more power in the prograde direction. This excess westward power leads to the observed super-rotation of the pattern. 
In the current paper we present new and independent evidence in support of these findings.
We note that the reality of the wave-like properties of supergranulation was questioned by \citet{rast_2004} and \citet{hathaway_2006}.

To further characterize the supergranulation phenomena observed by \citeauthor{gizon_2003} and \citeauthor{schou_2003}, here we analyze Dopplergrams and intensity images from the Helioseismic and Magnetic Imager \citep[HMI,][]{schou_2012} onboard the Solar Dynamics Observatory \citep[SDO,][]{pesnell_2012} using time-distance helioseismology (TD) and local correlation tracking (LCT), two independent techniques that are known to deliver consistent results for horizontal flow measurements near the solar surface \citep[e.g.,][]{langfellner_2015}. We then relate the flow evolution to the evolution of the magnetic field to study their interplay. To that end, we compare magnetograms measured by HMI with simple simulations, where corks represent the magnetic elements.

After introducing the observations, in this paper we show the signature of supergranulation in power spectra of the travel-time and LCT divergence maps on an extended wavenumber interval and study how this signature is influenced by the tracking rate. Then we present the horizontal flow evolution of the average supergranule over a range of spatial scales. In addition, we compare and discuss the results of the observed and simulated magnetic field. Finally, we draw conclusions for the dynamics of the supergranulation.

%%%%%%%%%%%%%%%%%%%%%%%%%%%%%%%%%%%%%%%%%%%%%%%%%%%%%%%%%%%%%%%

\section{Observations and data processing}  \label{sect_observations}

We used one year of SDO/HMI Dopplergrams and intensity maps in the period from May~2010 through April~2011. We tracked regions of size ${\sim}180 \times 180~$Mm$^2$ at the equator for 11~days, following those regions from about $70\degr$ east to $70\degr$ west. For the tracking rate, we used two options: a rate consistent with the rotation rate at the equator from \citet{snodgrass_1984}, and a rate faster than the \citeauthor{snodgrass_1984} rate by 60~m~s$^{-1}$ (in the rest of the paper dubbed ``fast tracking''). The fast rate corresponds to the supergranulation pattern rotation rate at the equator as measured by \citet{meunier_2007}, but is slower than the values obtained by \citet{duvall_1980} and \citet{snodgrass_1990}.
We remapped the regions using Postel's projection and a pixel size of $0.348~$Mm at a cadence of 45~seconds.

We then divided the 11-day tracked and mapped datacubes into 33 non-overlapping segments of 8~hour length each. Segment~17 thus crosses the central meridian. For each Dopplergram segment, we selected the f~modes by applying the ridge filter described in \citet{langfellner_2015}. We then computed point-to-annulus wave travel times \citep{duvall_1996} with an annulus radius of 10~Mm, using the method of \citet{gizon_2004}. These travel times, measured as the difference between the inward and outward travel times, are sensitive to the divergence of the horizontal flows near the solar surface.
For comparison, we also computed the horizontal flow divergence from local correlation tracking (LCT) of granules \citep{november_1988}. We first obtained the horizontal flow components using the code described in \citet{langfellner_2015} and then computed the divergence using Savitzky-Golay filters \citep{savitzky_1964}.

Following the procedures from our previous work \citep{langfellner_2015}, we constructed the average supergranule. To minimize edge effects, we cut away the outer 10~Mm in the travel-time maps and then removed the map-averaged travel time. We also removed the noise on scales $kR_\odot > 300$ by applying a low-pass filter in Fourier space.
In these processed travel-time maps, we identified the positions of both supergranular outflow and inflow centers, only using locations that are at least 18~Mm away from the map edges. We then co-aligned both the processed travel-time maps and the LCT horizontal divergence maps at these positions, always using the coordinates obtained for the segment in which the supergranules cross the central meridian (segment~17). This will allow us to study the evolution of the flows in the average supergranule near the solar surface.

To study the magnetic field in the context of the supergranular flows, we also obtained HMI line-of-sight magnetograms for the same times and regions as the Dopplergrams and intensity images. We averaged the magnetograms over 8~hours to reduce fluctuations on short timescales and took the absolute value to avoid cancellation of opposite polarities in the further processing. We then co-aligned these processed magnetograms at the supergranular positions.

%%%%%%%%%%%%%%%%%%%%%%%%%%%%%%%%%%%%%%%%%%%%%%%%%%%%%%%%%%%%%%%

\section{Supergranular power spectrum}  \label{sect_power}
\subsection{Dependence on tracking rate}
Using time-distance helioseismology and direct Doppler imaging, \citet{gizon_2003} and \citet{schou_2003} found that supergranulation exhibits a wave-like behavior. If studied at the equator, the wave power shows an excess power for waves travelling in the west (prograde) direction. \citet{rast_2004} claimed that these observations do not necessarily imply the existence of waves but could rather be artifacts arising from a tracking rate that is too slow.

To check this, we studied the power spectra of the divergence signal for two different tracking rates, one that is $0.028930~\mu\text{rad}~\text{s}^{-1}$ (about $20~$m~s$^{-1}$) slower \citep[the][tracking rate]{snodgrass_1984} and the other $0.057139~\mu\text{rad}~\text{s}^{-1}$
(about $40~$m~s$^{-1}$) faster (the ``fast'' tracking rate) than the Carrington rotation rate applied by \citet{gizon_2003}.
For each 11-day datacube, we collated the 33~travel time and LCT divergence maps in three-dimensional arrays and Fourier-transformed these arrays to obtain the power spectra.

\begin{figure*}[h]
\sidecaption
\parbox{0.7\hsize}{
\hspace{0.06\hsize} \framebox{\Large{Snodgrass tracking}}  \hspace{0.22\hsize}  \framebox{\Large{Fast tracking}} \\
\includegraphics[width=\hsize]{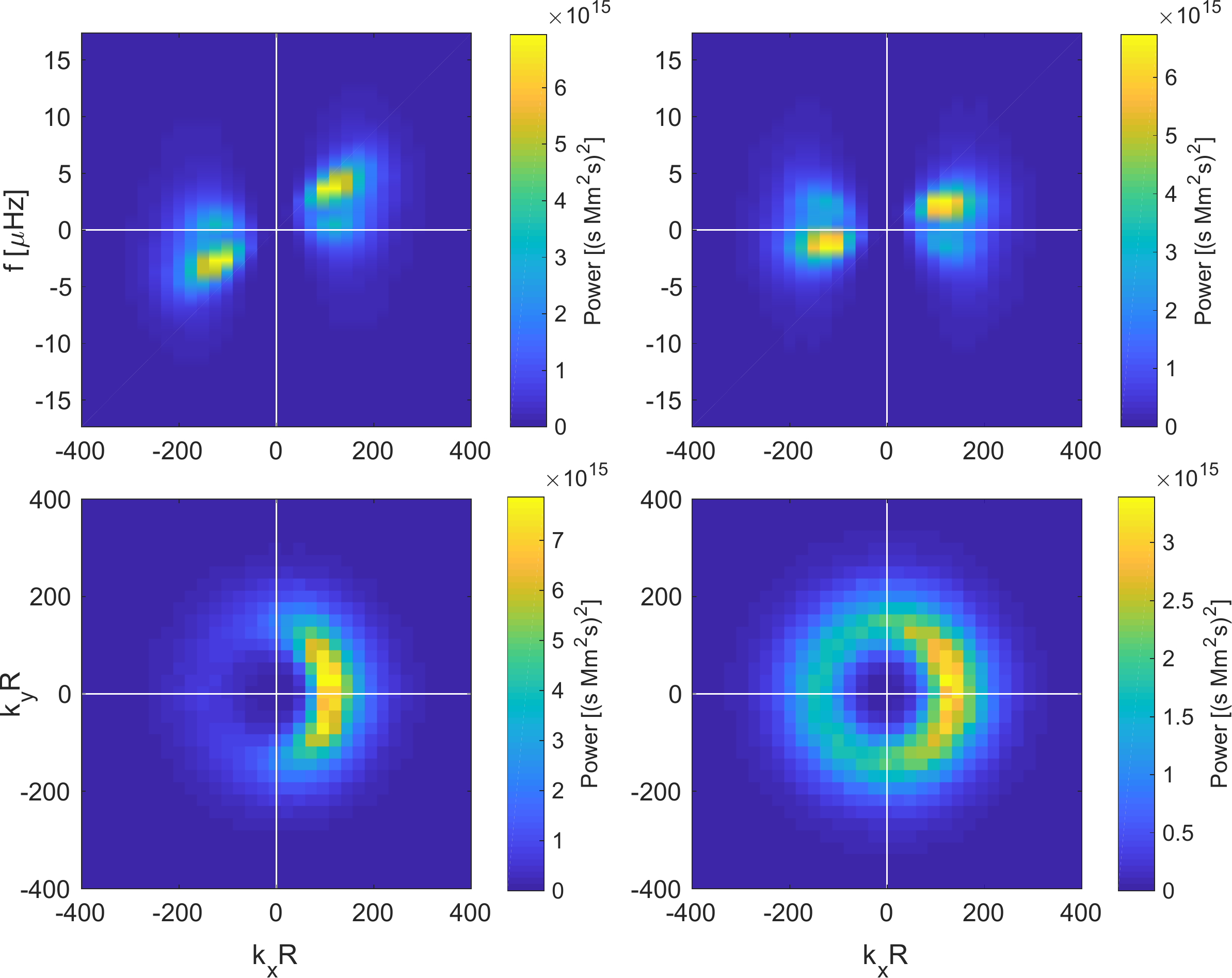}
}
\caption{Supergranular power spectrum of the divergence signal at the solar equator using time-distance helioseismology. \textit{Top row:} at $k_y=0$. The left panel is for a tracking rate set to \citet{snodgrass_1984} rotation rate (about 20~m~s$^{-1}$ slower than Carrington). The right panel is for a tracking rate that is faster by about 60~m~s$^{-1}$. \textit{Bottom row:} Cut through $k$ plane at frequency $3.2~\mu$Hz.}
\label{fig_sg-power}
    \end{figure*}

\begin{figure*}[h]
\sidecaption
\includegraphics[width=0.7\hsize]{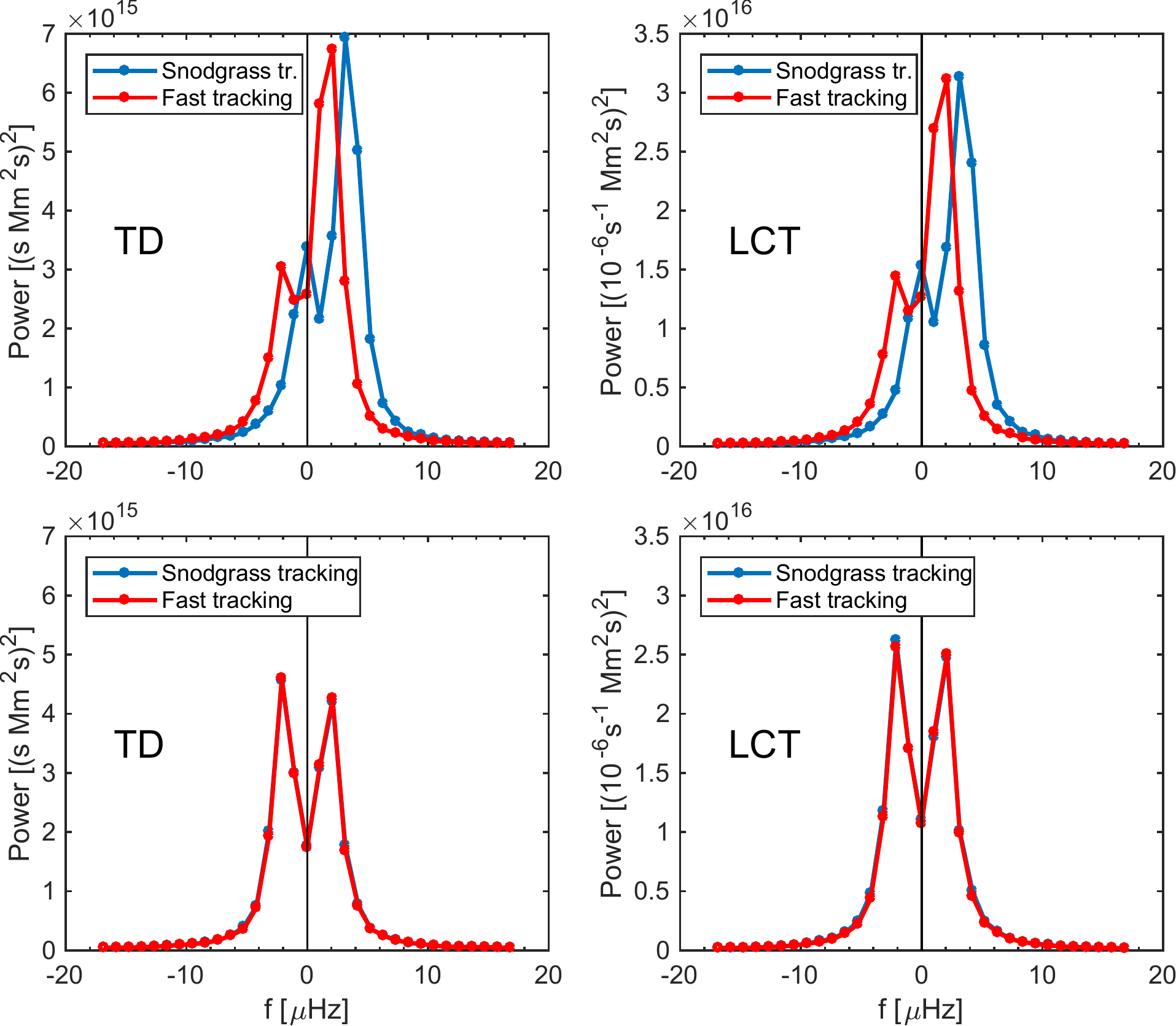}
\caption{Supergranular power spectrum of the divergence signal at the solar equator for both tracking rates, cut at $kR_\odot \approx 123$, using one year of HMI data: time-distance helioseismology (left) and local correlation tracking (right). \textit{Top row:} East-west direction. \textit{Bottom row:} North-south direction.}
\label{fig_sg-power_cut}
    \end{figure*}

Figure~\ref{fig_sg-power} shows various cuts through the three-dimensional power spectra for the travel-time cubes at the solar equator, using the two tracking rates and TD. The power is concentrated in a ring at supergranular scales ($kR_\odot \sim 100$), with most power in the prograde direction. In the east-west direction, cutting at $k_x \approx 120/R_\odot$ and $k_y=0$ (see Fig.~\ref{fig_sg-power_cut}, upper-left panel), two peaks are visible for the \citeauthor{snodgrass_1984} tracking rate. The stronger peak is located at about $3~\mu$Hz, the weaker one at ${\sim}0~\mu$Hz. For the faster tracking rate, the power appears shifted to the lower frequencies, so that the two peaks have equal distance from zero frequency (${\sim} {\pm} 2~\mu$Hz), but otherwise the power distribution is very similar to the \citeauthor{snodgrass_1984} tracking rate. For LCT (upper-right panel), the power distribution is almost indistinguishable from the time-distance case. The frequency shift of the peaks in the east-west direction can be explained by the Doppler shift $\Delta\omega/2\pi = ku_x \approx 1.7~\mu$Hz due to the tracking rate difference $u_x \approx 60$~m~s$^{-1}$. In the north-south direction ($k_x=0$, $k_y \approx 120/R_\odot$, lower panels), the two peaks are roughly equal in height and are located at ${\sim} {\pm} 2~\mu$Hz, irrespective of the tracking rate. Again, TD and LCT agree well.

%______________________________________________ SG power spectra

%______________________________________________ SG power spectra cuts: kxR=120, ky=0

\subsection{Fitting a parametric wave model}
The power distribution and amplitude ratios of the peaks are in agreement with \citet[][cf.~Fig.~1 therein]{gizon_2003}.
To extract wave properties like the dispersion relation, we fit the power, separately for each wavenumber $k$, using a parametric wave model that builds on the model presented in \citet{gizon_2004a}. The model consists of a sum of two Lorentzians, centered at frequencies $\pm\omega_0$, whose amplitudes depend on the azimuth $\psi$. The wavevector $\textbf{k} = (k_x, k_y)$ is parameterized as $\textbf{k}= (k\cos\psi,k\sin\psi)$. In addition, the model takes into account the Doppler shift $\textbf{k}\cdot\textbf{u}$ from a horizontal flow $\textbf{u} = (u_x,u_y)$, a finite lifetime $t_\text{life}=1/\gamma$, and a background term that is independent of frequency. The full model is given by
\begin{align}
 P(\textbf{k},\omega) &= \frac{F(\textbf{k},\omega) + F(-\textbf{k},-\omega)}{2} + B_0(k) + B_2(k)\cos(2\psi - \alpha), \\
 F(\textbf{k},\omega) &= \frac{A_0(k) + A_1(k) \cos(\psi - \psi_\text{max})+ A_2(k) \cos(2\psi - \alpha)}{1+[\omega-\omega_0(k)-\textbf{k}\cdot\textbf{u}(k)]^2/\gamma(k)^2},
\end{align}
where $\psi_\text{max}$ is the direction of maximum power, $A_1/A_0$ gives the power anisotropy, and the $A_2$ term describes spatial variations with respect to an angle $\alpha$ that can arise from, e.g., instrumental astigmatism. The coefficients $B_0$ and $B_2$ describe the background term.

%______________________________________________ tau^oi and LCT power: fits
   \begin{figure*}[h]
\sidecaption
\includegraphics[width=0.7\hsize]{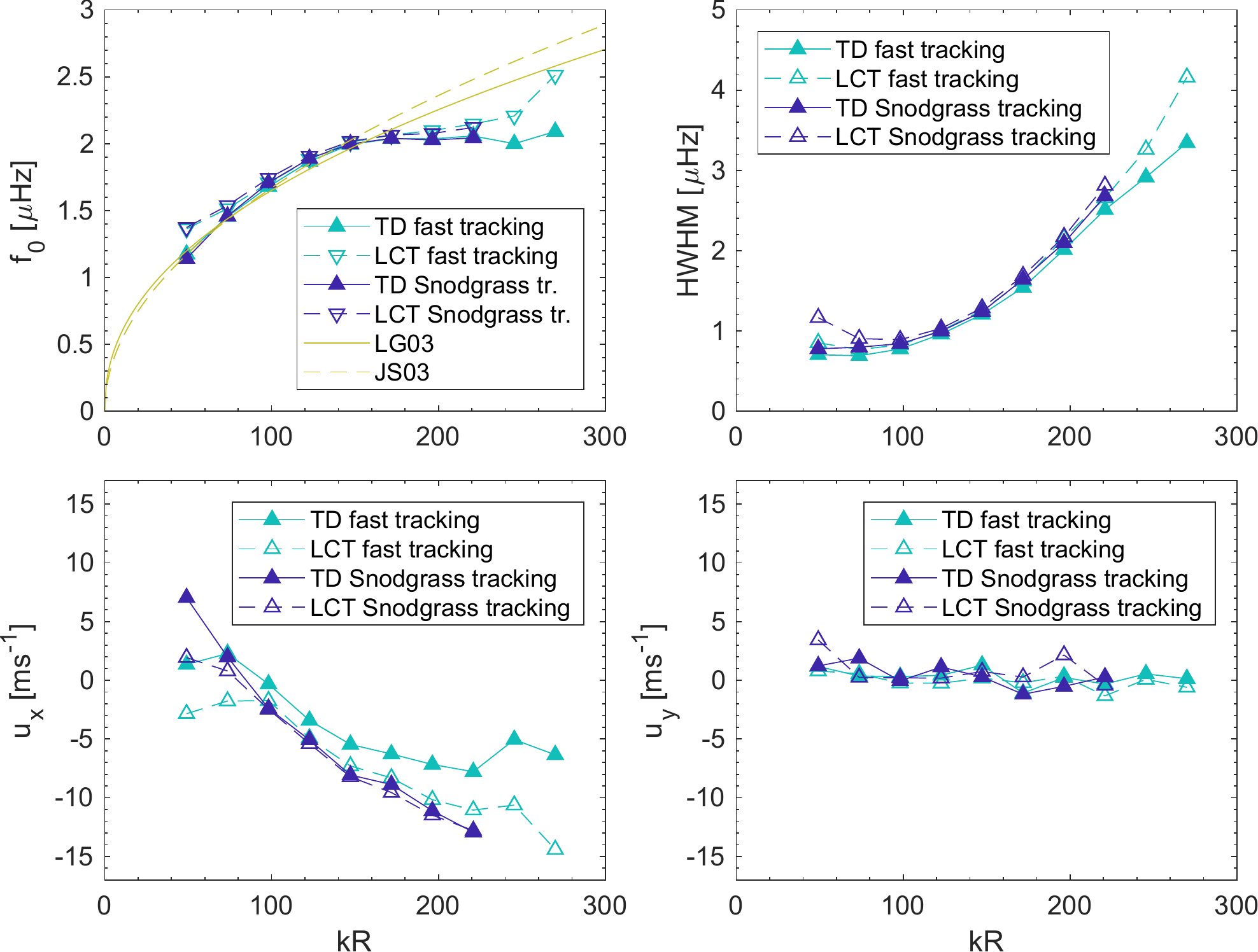}
\caption{Parameters from Lorentzian fits of time-distance and LCT divergence power spectra for \citet{snodgrass_1984} and fast tracking rates. \textit{Top row:} Dispersion relation and half-width at half maximum (HWHM). \textit{Bottom row:} Horizontal velocity components of background flow. The tracking rate difference of $60~$m~s$^{-1}$ has been subtracted from the $u_x$ values for \citet{snodgrass_1984} tracking to facilitate the comparison with fast tracking. Values for $kR_\odot > 230$ (\citeauthor{snodgrass_1984} tracking) and $kR_\odot > 271$ (fast tracking) are omitted, as the fits are not stable in that regime.}
\label{fig_power-fit-dispersion}
    \end{figure*}

The observed and fitted power as a function of azimuth and frequency is shown for $kR_\odot \approx 123$ in Figs.~\ref{fig_sg-power_azimuth_snodgrass} and \ref{fig_sg-power_azimuth_fast} in the Appendix. For the \citeauthor{snodgrass_1984} tracking rate, the bands of strong power are located at frequencies that depend on azimuth, consistent with a Doppler shift $\Delta\omega = \textbf{k}\cdot\textbf{u} = ku_x \text{cos}\psi$ due to a flow $u_x \approx 60~$m~s$^{-1}$. Adjusting the tracking rate, as we did in the fast tracking case, removes this Doppler shift and thus the $\text{cos} \psi$ dependence of the frequencies where the power has a maximum, but leaves the power spectrum otherwise largely unaffected (see Fig.~\ref{fig_sg-power_azimuth_fast}).

%______________________________________________ tau^oi and LCT evolution: fits
   \begin{figure*}[h]
\sidecaption
\includegraphics[page=1,width=0.7\hsize]{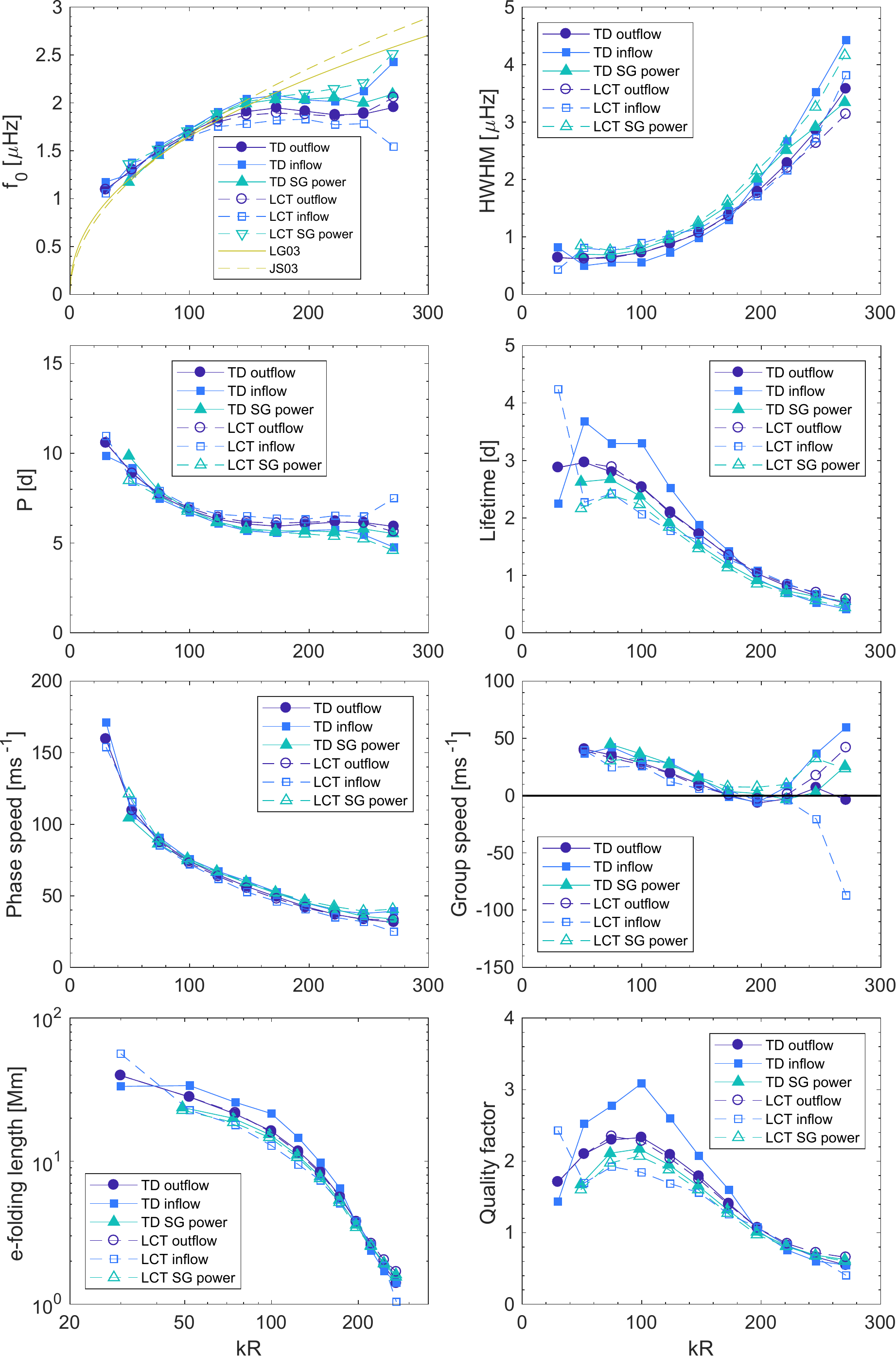}
\caption{Parameters extracted from Lorentzian fits of divergence power spectra and the average convection cell evolution (see Sect.~\ref{sect_flows-scales}) for fast tracking rate at the solar equator, as a function of spatial scale. Values for $kR_\odot > 271$ are omitted as the fits are not stable in that regime. For $f_0$ the dispersion relations from \citet{gizon_2003} and \citet{schou_2003} are overplotted for comparison.}
\label{fig_fit-parameters}
    \end{figure*}

Figure~\ref{fig_power-fit-dispersion} shows the dispersion relation (with $f_0 =\omega_0/2\pi$) that we obtained through the fits of the power spectra. The curves for TD and LCT agree well and do not show any dependence on the tracking rate. Moreover, the dispersion relation that we measure is consistent with the empirical power law estimate obtained by \citet{gizon_2003} and \citet{schou_2003} for $kR_\odot<150$. At larger wavenumbers, our measured frequencies are systematically below the power law estimates. However, in the interval $150<kR_\odot<200$, the values obtained by \citet{gizon_2003} and \citet{schou_2003} are not well described by their power laws either. Indeed, their frequency values are ${\sim}2.0~\mu$Hz for these wavenumbers, which is consistent with our measurements. The regime at $kR_\odot>200$ was not explored previously. Our data show a continuation of the flat dispersion relation until the fits start to diverge at $kR_\odot\sim270$.

The half-width at half maximum (HWHM), $\gamma/2\pi$, of the Lorentzians is independent of the tracking rate as well. It is roughly constant for $kR_\odot<100$ and then starts to increase at an accelerating rate for larger wavenumbers. At $kR_\odot \approx 200$, the HWHM becomes larger than $f_0$. This behavior agrees well with \citet{gizon_2003} and \citet{schou_2003}, confirming their findings.

The bottom row of Fig.~\ref{fig_power-fit-dispersion} shows that the flow speed $u_x$ is not constant, but decreases with $kR_\odot$ for $kR_\odot<240$. This could be a sign that different scales of supergranules are affected differently by the radial differential rotation in the near-surface shear layer, and thus probe different depths. The North-South flow component $u_y$, on the other hand, is consistent with zero for all scales and both tracking rates, as expected from meridional circulation at the solar equator.

For the fast tracking, Fig.~\ref{fig_fit-parameters} shows some selected parameters derived from the dispersion relation and ridge width of the power spectra, together with values derived from the analysis of the average supergranule that is described in Sect.~\ref{sect_flows-scales}. The period $P = 1/f_0$ varies from 10~days at $kR_\odot \approx 40$ to 6~days at smaller wavenumbers ($kR_\odot > 120$). The lifetime, $t_\text{life} = 1/\gamma$, decreases with wavenumber, from about 3~days at the largest scale to half a day at $kR_\odot \approx 250$. The phase speed, $c_\text{ph} = \omega_0/k$, drops quickly from about 150~m~s$^{-1}$ at the largest scale to 60~m~s$^{-1}$ at $kR_\odot = 120$, and then decreases slowly to 30~m~s$^{-1}$ at $kR_\odot = 250$. The group speed $c_\text{gr} = \partial \omega_0/\partial k$, on the other hand, shows a peculiar decline from about 40~m~s$^{-1}$ at $kR_\odot \approx 40$ to zero at $kR_\odot = 200$. This coincides with a drop of the quality factor ($f_0/\text{HWHM}$) below one. The highest quality factor is reached at $kR_\odot \sim 90$. Finally, the e-folding length $c_\text{ph} t_\text{life}$ is monotonically decreasing with wavenumber. There is a conspicuous change of slope in the log-log plot at $kR_\odot \approx 120$, the scale of strongest power for supergranulation. At larger scales, the e-folding length increases more slowly than the wavelength. We might speculate that this is connected to a convective regime change at this scale, as proposed by, e.g., \citet{cossette_2016} and \citet{rincon_2017}.

\subsection{Distinguishing wavenumber regimes}
When considering the observations of the supergranular power spectrum at a slightly higher level of abstraction, the following wavenumber intervals can be distinguished:
\begin{enumerate}
\item $kR_\odot\lesssim 120$: The dispersion relation can be described by a squareroot function and there are two clearly distinguishable peaks in frequency cuts of the power spectrum (see, e.g., Fig.~\ref{fig_sg-power_cut}).
\item $120 \lesssim kR_\odot\lesssim 200$: The dispersion relation is flat ($f_0 \approx 2~\mu$Hz) and there is no more visible minimum between the two peaks in the east-west direction -- this is mostly because the power in the east direction is much smaller than in the west direction.
\item $200 \lesssim kR_\odot\lesssim 270$: The quality factor has fallen below one (the HWHM of the ridge exceeds its central frequency $f_0$).
\item $kR_\odot\gtrsim 270$: Even in the north-south direction, where the two peaks are of equal strength, there is no more visible minimum between the two peaks. Analogous to the Rayleigh criterion in optics, this provides an estimate for the resolution limit. This is supported by the fact that the Lorentzian fits are diverging or completely failing in this regime, so no meaningful dispersion relation is extractable beyond $kR_\odot \sim 270$.
\end{enumerate}

%%%%%%%%%%%%%%%%%%%%%%%%%%%%%%%%%%%%%%%%%%%%%%%%%%%%%%%%%%%%%%%

\section{Supergranular flow evolution}
\subsection{The average supergranule}
We now analyze the evolution of the intermediate-scale convection in real space by measuring the near-surface flows of the average supergranule at a reference time and following it up to five days into the future and past.

\begin{figure*}[h]
\centering
\includegraphics[width=\hsize]{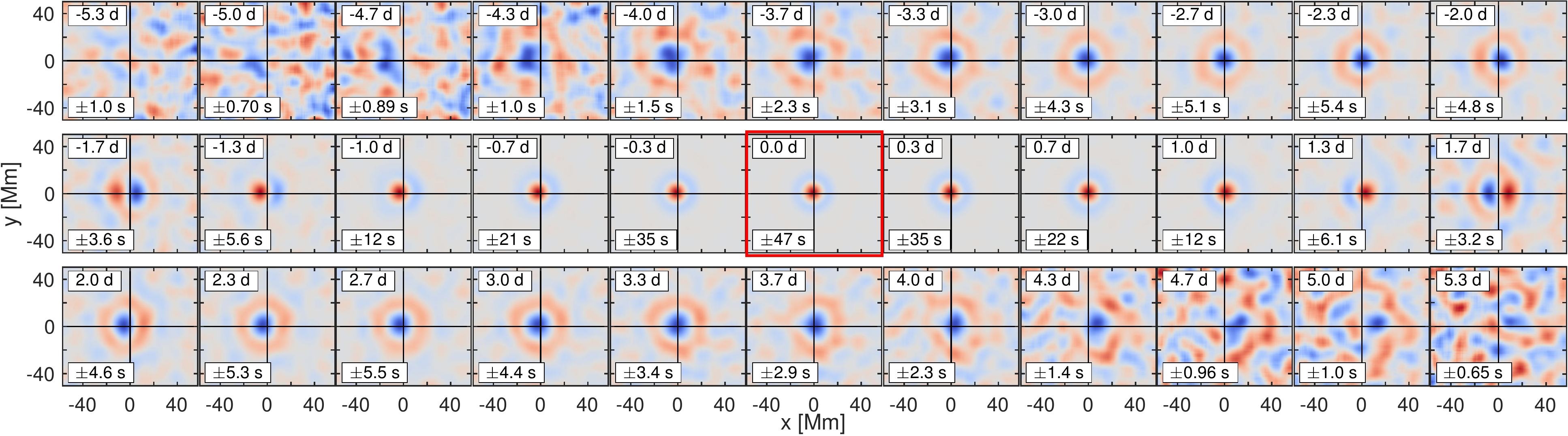}  \\
\includegraphics[width=\hsize]{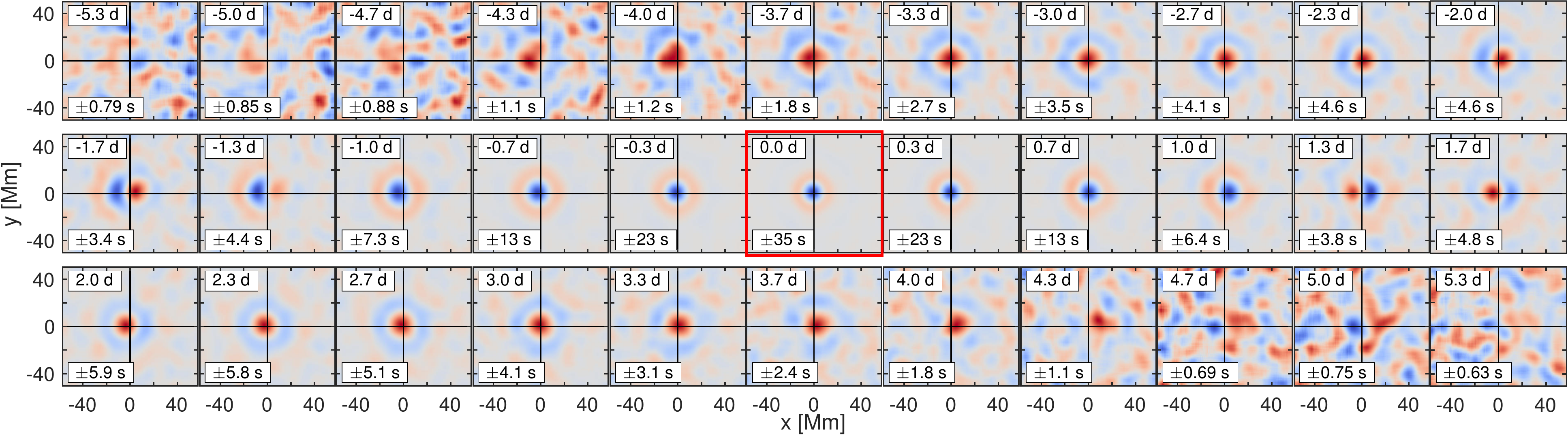}
\caption{Evolution of f-mode-filtered inward$-$outward travel times for the average supergranule. \textit{Top:} Outflow. \textit{Bottom:} Inflow. The frames, from top left to bottom right, span a total of eleven days in non-overlapping steps of eight hours. The central frame (red outline) is for the reference time, $\Delta t=0$. Note that the color scale is different for each individual frame; saturation is reached at the maximum absolute value (given by the lower label boxes) of the respective frame. The color map is symmetrized around zero; outflows are red and inflows are blue.}
\label{fig_tauoi}
    \end{figure*}

Figure~\ref{fig_tauoi} shows the evolution of inward$-$outward travel times for the average supergranular outflows and inflows.\footnote{Movies of the evolution for TD and LCT can be found on-line. See Appendix~\ref{sect_movies} for additional information.}
The travel-time maps show a complex evolution with an oscillatory component. At the reference time, $\Delta t = 0$, the central outflow region is surrounded by a ring of inflows. The outflow region drifts westward for $\Delta t > 0$, while the eastern part of the inflow ring follows in the same direction and gains strength. After a short dipolar stage, where both components are roughly equally strong ($\Delta t \sim 1.7~$d), the inflow region moves further west into the center. After about three days, the pattern seen at the reference time has reversed: a central inflow region is now surrounded by a ring of outflows. However, the size of the structure has increased; the ring diameter is roughly 50~Mm compared to about 30~Mm at the reference time. For larger time lags ($\Delta t > 4~$d), the signal mostly vanishes beneath the noise background. Yet it can be seen that at $\Delta t = 5.3~$d the sign at the origin reappears and switches back to the sign at reference time. Before the reference time ($\Delta t < 0$), a similar behavior is observed as for positive $\Delta t$, albeit with the time axis reversed and the east-west direction flipped. At times $\Delta t \sim -3~$d and $\Delta t \sim +3~$d, the patterns are very similar -- the supergranulation pattern thus appears to undergo an oscillation with a period of ${\sim}6$~days.

%______________________________________________ tau^oi evolution for avg. SG outflow and inflow

For the average inflow at reference time, the evolution is very similar to the outflow but with the sign of the travel times reversed. Small differences to the average outflow can be noticed, which manifest as a generally smaller travel-time amplitude and a slightly shorter oscillation period.

\begin{figure*}[h]
\centering
\includegraphics[width=\hsize]{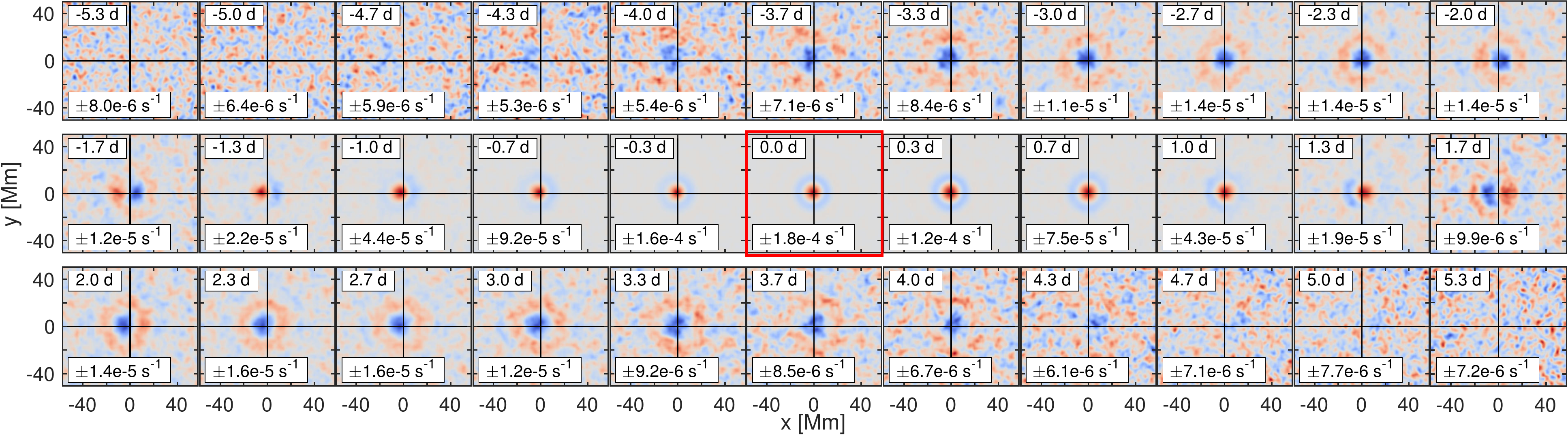}
\caption{Evolution of LCT $\text{div}_h$ for the average supergranular outflow (cf.~top of Fig.~\ref{fig_tauoi}).}
\label{fig_LCTdiv}
    \end{figure*}

In LCT divergence maps (Fig.~\ref{fig_LCTdiv}, using the same supergranule positions as before), the evolution of the average supergranule shows the same characteristics as for the travel-time maps. Outflows change into inflows and vice versa in about three days; the ring diameter increases away from the reference time. The patterns in the LCT divergence maps are equal to the travel-time maps down to the smallest scales present in the (low-pass-filtered) inward$-$outward travel-time maps. There is, however, a slight asymmetry in time; for the average outflow, the divergence is stronger before the reference time, for the average inflow (not shown), it is stronger after the reference time. The reason for this temporal asymmetry is unclear; among the possible causes are the different depth sensitivities of f-mode TD and granule LCT or a systematic influence of the magnetic field on the divergence measurements.

%______________________________________________ LCT div evolution for avg. SG outflow

To summarize, the supergranular evolution at the equator is both symmetric in time and the east-west direction; inflows and outflows turn into each other and can essentially be regarded as phase-shifted versions of one another.
In other words, new supergranules preferentially form in old inflows. This is consistent with the findings by \citet{shine_2000} who observed several individual supergranules forming in LCT divergence maps obtained from Michelson Doppler Imager \citep{scherrer_1995} intensity images.
The sign reversal of the horizontal divergence after about two days was also observed by \citet{greer_2016} using helioseismic ring-diagram analysis (see Figs.~7 and 8a therein). The authors measured a super-rotation velocity of the supergranular pattern of about 50~m~s$^{-1}$ compared to the Carrington rotation rate, which is slightly faster than our fast tracking rate (40~m~s$^{-1}$).

The oscillatory behavior does not depend on the tracking rate. Figure~\ref{fig_tauoi_tracking} in the Appendix shows the evolution of the average supergranular outflow for the \citet{snodgrass_1984} tracking rate, using time-distance helioseismology. The visible patterns are almost indistinguishable from the $60~$m~s$^{-1}$ faster supergranule pattern rotation rate (Fig.~\ref{fig_tauoi}), except for an east-west drift of the pattern in the \citeauthor{snodgrass_1984} case that is consistent with the difference in the tracking rates. This clearly contradicts the claims by \citet{rast_2004} and \citet{hathaway_2006} that the measured oscillation is an artifact due to tracking rate or projection effect issues.

In the remainder of this paper, we will only consider the faster tracking rate. This will facilitate the analysis of the supergranular pattern evolution over the course of days, as the envelope maximum of the pattern does not drift in the east-west direction.

\subsection{Dependence on spatial scale} \label{sect_flows-scales}
\begin{figure*}[h]
\centering
\includegraphics[width=\textwidth]{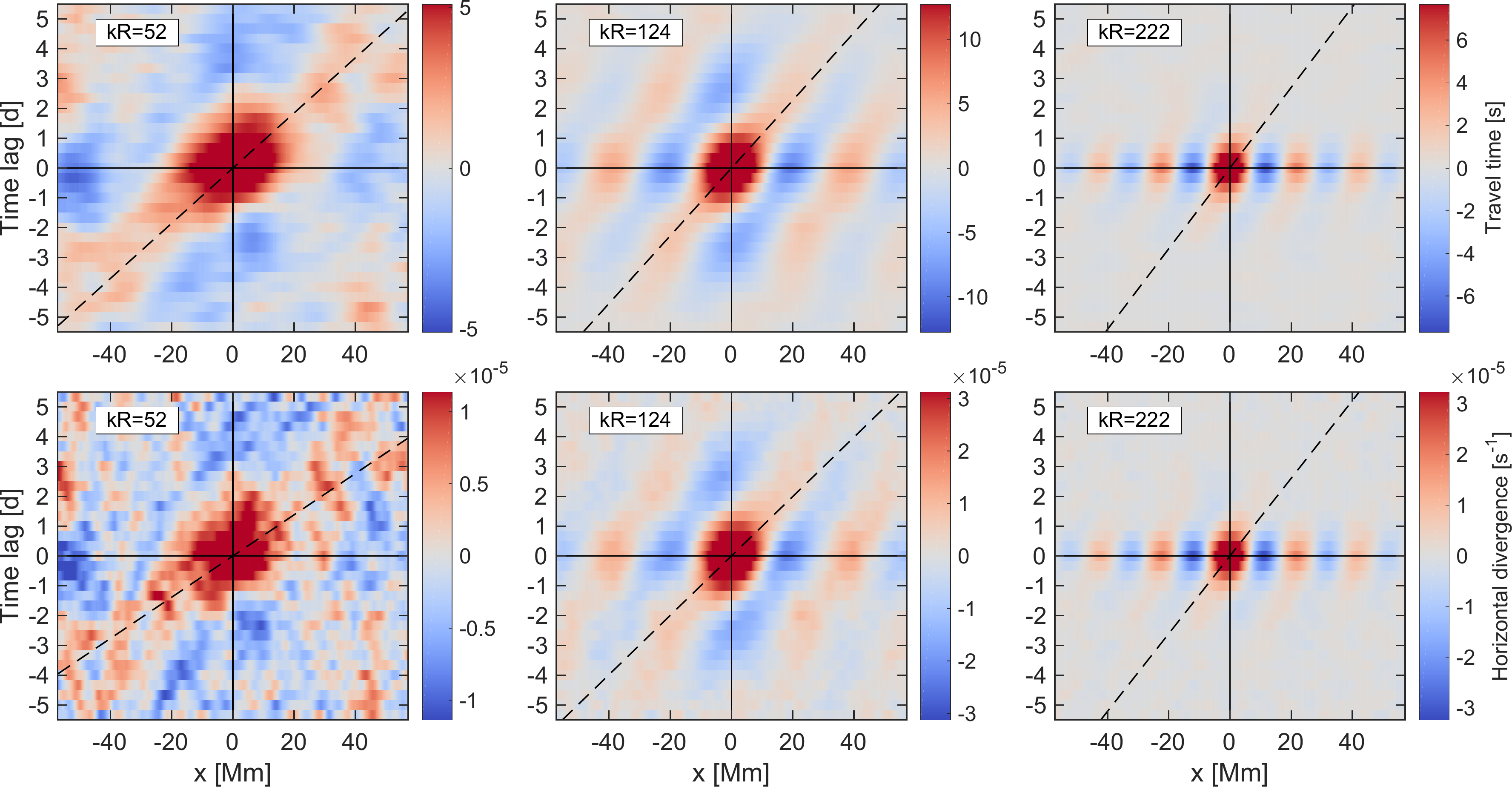}
\caption{Evolution for different scales: space-time images. Average over $y$ band (10 Mm) for the average outflow. \textit{Top row:} TD. \textit{Bottom row:} LCT. The dashed lines give the phase speed from the dispersion relation as obtained from the power spectrum. Outflows are red and inflows are blue. To increase visibility, the color maps are saturated at half the maximum value.}
\label{fig_evolution-yavg}
    \end{figure*}

In order to study the change in size of the supergranular evolution pattern and to make a connection to the findings from the power spectra,
we investigated the evolution of the average cell separately for a set of narrow scale intervals, spanning the overall scale range from about 15~Mm to 80~Mm.
To do so, we band-pass-filtered the travel-time maps in Fourier space to only select supergranular cells in a narrow scale range before applying the image segmentation algorithm to identify the positions of the cells. For each band-pass filter, we then co-aligned the TD and LCT maps (low-pass-filtered and not filtered, respectively) at these positions as before. The results for convective outflows at three selected scales are shown as space-time plots in Fig.~\ref{fig_evolution-yavg}.

The oscillatory behavior -- outflows turn into inflows and vice versa -- can be observed at all the scales that we analyzed, with good agreement between TD and LCT. When comparing the scales by eye, two things stand out: The oscillation period decreases with higher wavenumber (roughly from 8~d to 4~d) and the lifetime of the pattern decreases (at $kR_\odot = 52$, the pattern is visible in all panels, whereas at $kR_\odot = 222$, it disappears by $|\Delta t| \approx 3~$d).

These observations are in line with the dispersion relation and linewidth that we obtained from the power spectra as visualized in Sect.~\ref{sect_power}, Fig.~\ref{fig_fit-parameters}.
Whereas at small wavenumbers the travel times fall off slowly at larger time lags, at larger wavenumbers the travel times are increasingly concentrated in thin strips around $\Delta t=0$. The decreasing phase speed with larger wavenumbers manifests itself as a steeper slope of the red and blue diagonal stripes.

%______________________________________________ tau^oi evolution: y-band average outflow (space-time image)

\begin{figure*}[h]
\sidecaption
\includegraphics[width=0.7\hsize]{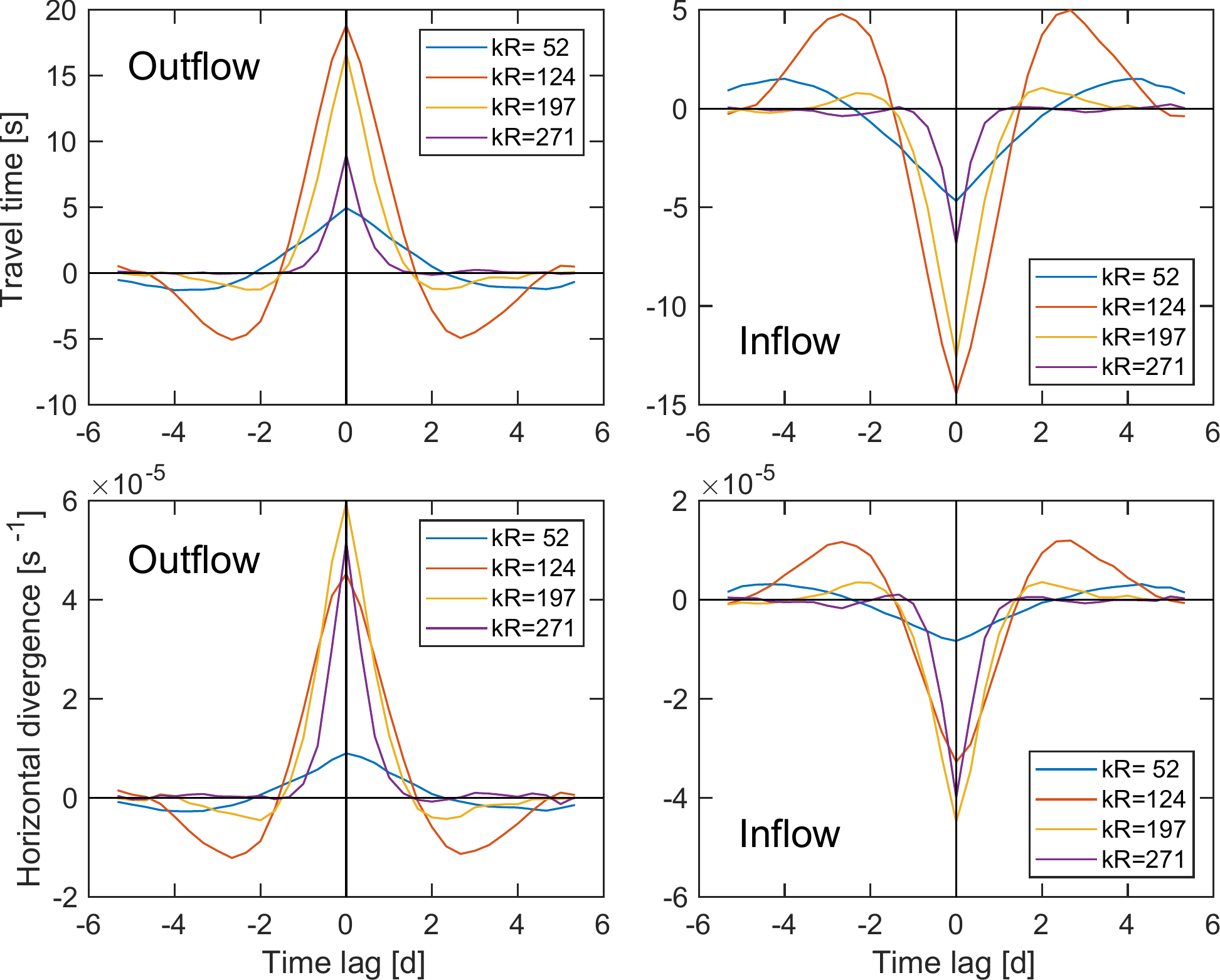}
\caption{Evolution of supergranular cells on different scales: Average over circular area with radius $(2\pi R_\odot/k)/4$ around the origin. \textit{Top row:} TD outflow and inflow (sign reversed). \textit{Bottom row:} LCT outflow and inflow.}
\label{fig_evolution-fit}
    \end{figure*}

For a quantitative comparison of the supergranular evolution pattern and the power spectra, we studied the time dependence of the convective signal at the origin ($x=0$, $y=0$) and average over a small circular region to increase the signal-to-noise ratio. This is similar to studying the autocorrelation of travel-time and LCT divergence maps on a given spatial scale. The results are shown in Fig.~\ref{fig_evolution-fit}. Since the autocorrelation is the inverse Fourier transform of the power spectrum, we are motivated to fit the inverse temporal Fourier transform of a Lorentzian to the curves:
\begin{equation}
 f(k,\Delta t) = a(k) \cos(\omega_0(k)\Delta t) \exp(-\gamma(k) |\Delta t|),  \label{eq_lorentzian}
\end{equation}
where $\omega_0$ and $\gamma$ are as defined in Sect.~\ref{sect_power}.

The fit results are overplotted in Fig.~\ref{fig_fit-parameters} in Sect.~\ref{sect_power}. The parameters agree well with the fit parameters obtained from the power spectra. This means that we can identify the structures in the power spectra with the observed supergranular evolution pattern.

The spatial scale $kR_\odot = 271$ is the limiting case where the oscillations (corresponding to the cosine term in Eq.~\ref{eq_lorentzian}) are barely visible, so it is hard to extract a meaningful period beyond this scale. This is reflected by the low quality factor in Fig.~\ref{fig_fit-parameters}. On larger scales though, oscillations are clearly present -- lifetimes and periods can thus be measured simultaneously even if the periods are longer than the lifetimes (e-folding times).

Fig.~\ref{fig_evolution-fit} also shows that LCT is more sensitive to smaller scales than the TD travel times. For example, the LCT divergence amplitudes at zero time lag are larger at $kR_\odot = 271$ than at $kR_\odot = 124$, whereas the travel-time amplitudes at $kR_\odot = 271$ are only about half the amplitude at $kR_\odot = 124$.

%______________________________________________ tau^oi and LCT evolution: average over circle vs. time lag

Figures~\ref{fig_tauoi} and \ref{fig_LCTdiv} can now be better understood in light of the scale dependence: The average supergranule that was found without band-pass filtering is a power-weighted average over a range of spatial scales. The larger-scale flows have longer lifetimes, thus the shorter-lived smaller-scale flows decay more rapidly and the average scale increases with larger $|\Delta t|$. The average supergranular inflow apparently has a slightly shorter oscillation period (and a weaker divergence signal) than the outflow. A possible explanation is that slightly smaller scales (which might be concealed due to smoothing of the measurements) are selected on average for the inflows.

%%%%%%%%%%%%%%%%%%%%%%%%%%%%%%%%%%%%%%%%%%%%%%%%%%%%%%%%%%%%%%%

\section{Magnetic field evolution}

\subsection{Observations}
\begin{figure*}[h]
\centering
\includegraphics[width=\hsize]{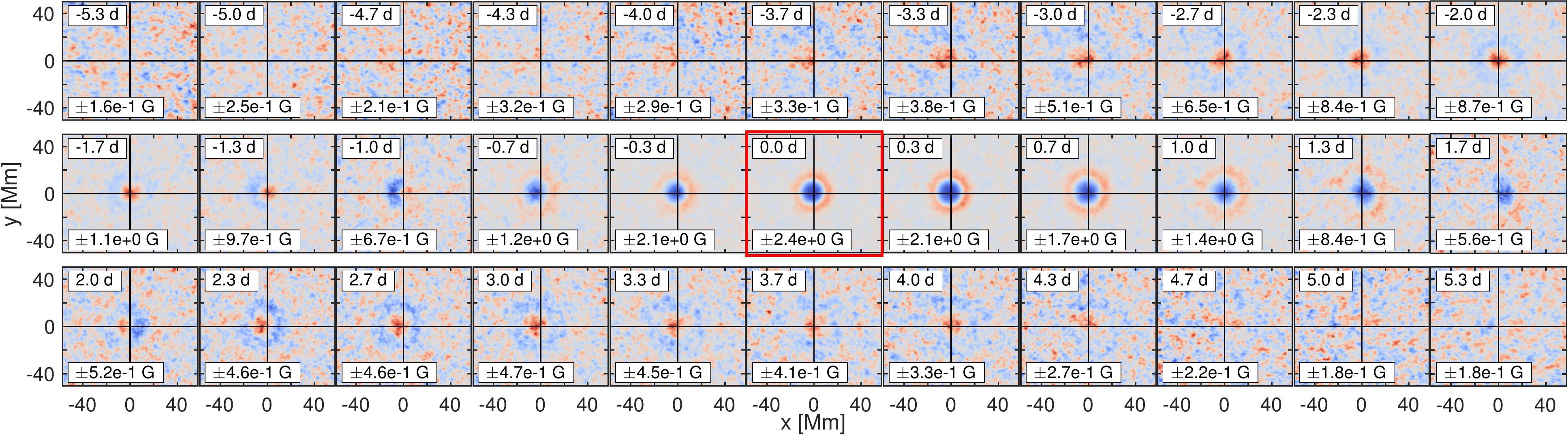}
\includegraphics[width=\hsize]{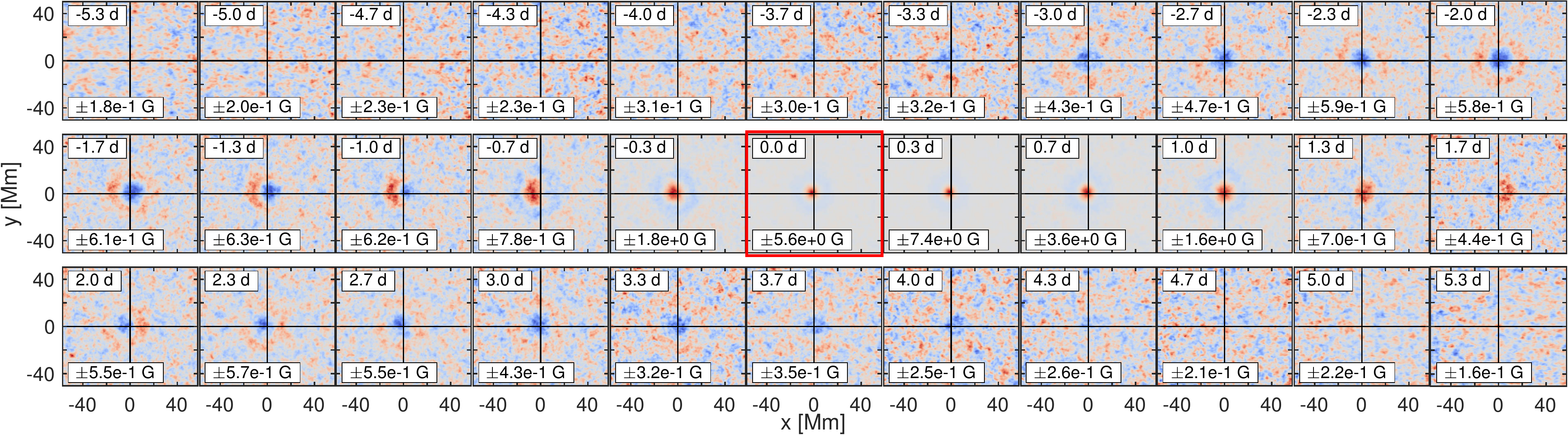}
\caption{Evolution of magnetic field (absolute line-of-sight field, mean subtracted) for the average supergranule. \textit{Top:} Outflow. \textit{Bottom:} Inflow. Blue is weaker than the average field strength and red is stronger than average.}
\label{fig_B}
    \end{figure*}

In addition to the horizontal flows, we studied the evolution of the line-of-sight magnetic field for the average supergranule. The data processing steps are outlined in Sect.~\ref{sect_observations}. Figure~\ref{fig_B} shows that the magnetic field follows the oscillations of the supergranular flows. When there is an average inflow, the magnetic field is concentrated and stronger than average. When there is an outflow, on the other hand, the magnetic field is diluted.
However, the evolution is not entirely symmetric in time with respect to the reference time ($\Delta t = 0$, time of strongest outflow/inflow). For example, the cleared zone in the average outflow is still clearly present at $\Delta t = 1.3~$d, but is only starting to form at $\Delta t = -1.3~$d.

For the average inflow, the maximum magnetic field strength is not reached at the reference time, but a few hours later. Assuming a smooth evolution, a temporal Fourier interpolation yields a time lag of about $6~$hours (see right panel in Fig.~\ref{fig_cork-timelag}).

%______________________________________________ B evolution for avg. SG outflow/inflow

\subsection{Cork simulation} \label{sect_corksim}
The observed time lag in the magnetograms raises the question if the magnetic field is merely passively advected by the supergranular flows or if it also plays an active part in shaping those flows. The latter was proposed by, e.g., \citet{crouch_2007}, who suggested that random clustering of magnetic elements could trigger downflows on supergranulation scales. Recently, \citet{chatterjee_2017} found evidence for a solar-cycle dependence of supergranular parameters (e.g., the cell size) in century-long \ion{Ca}{II}~K observation records.

To study how much the magnetic field of the average supergranule is influenced by the supergranular flows, we conducted a cork simulation, ignoring magnetic polarity. Magnetic elements were modeled as non-interacting floating corks with a finite lifetime $t_\text{life} = 16$~h that corresponds to the flux replacement time \citep{hagenaar_2003}. The corks are advected by supergranulation-scale flows and simultaneously perform a random walk owing to the statistical influence of smaller-scale convective flows \citep[as seen by][for the internetwork field]{orozco_2012,roudier_2016,agrawal_2018} with a turbulent diffusivity $\eta = 250~$km$^2$~s$^{-1}$ \citep{simon_1997}. Further correlations in the smaller-scale flows, e.g., the coherent flows on mesogranular scales that build up in trees of fragmenting granules, as described by \citet{roudier_2016}, were neglected. Details of the simulation are given in Appendix~\ref{sect_corksim-details}.

We then computed the cork density by counting the number of corks in each pixel (on the observation grid) for a given time step. To facilitate the comparison with the observed magnetic field, we smoothed the cork density temporally by applying an 8-hour boxcar moving average and spatially by convolution with a Gaussian of 0.7~Mm FWHM, similar to the HMI point-spread function \citep{yeo_2014}.

%______________________________________________ cork evolution for avg. SG outflow/inflow
   \begin{figure*}[h]
\centering
\includegraphics[width=\hsize]{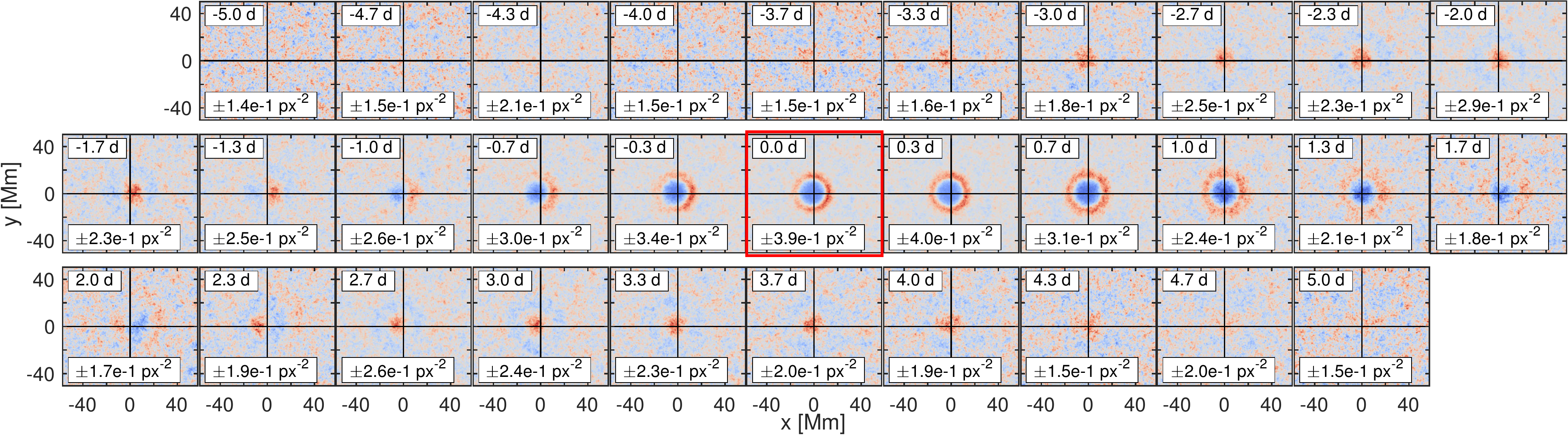}
\includegraphics[width=\hsize]{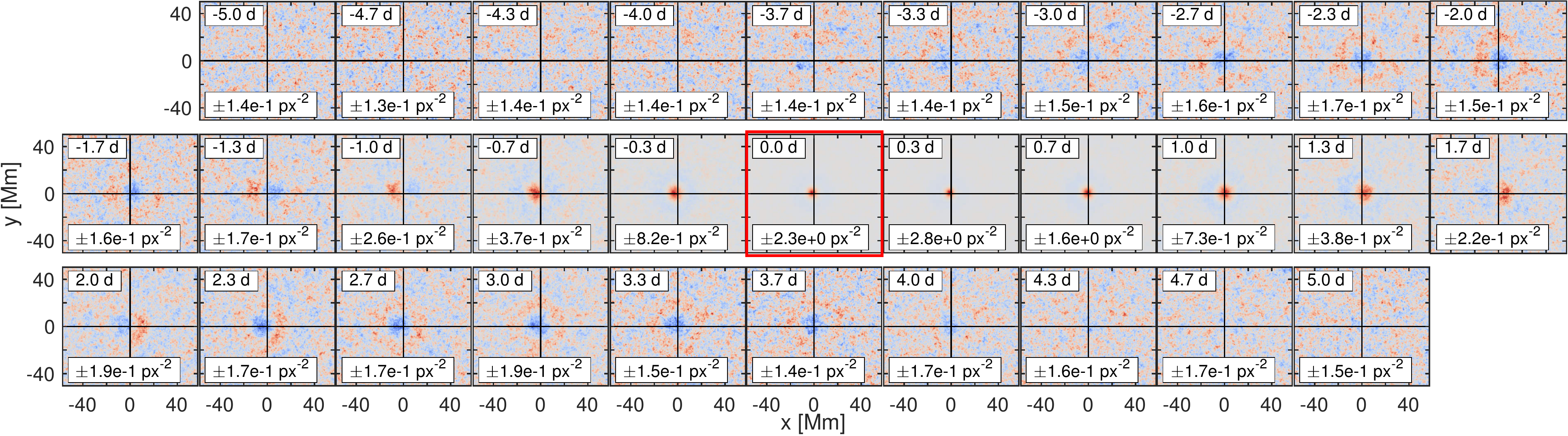}
\caption{Simulated evolution of cork density (mean subtracted) for the average supergranule. \textit{Top:} Outflow. \textit{Bottom:} Inflow. Blue is weaker than the average cork density, red is stronger than average. The cork density is taken from a simulation run with $\eta = 250~$km$^2$~s$^{-1}$ and $t_\text{life} = 16$~h (run \#5).}
\label{fig_cork}
    \end{figure*}

\begin{figure*}[h]
\sidecaption
\includegraphics[width=0.7\hsize]{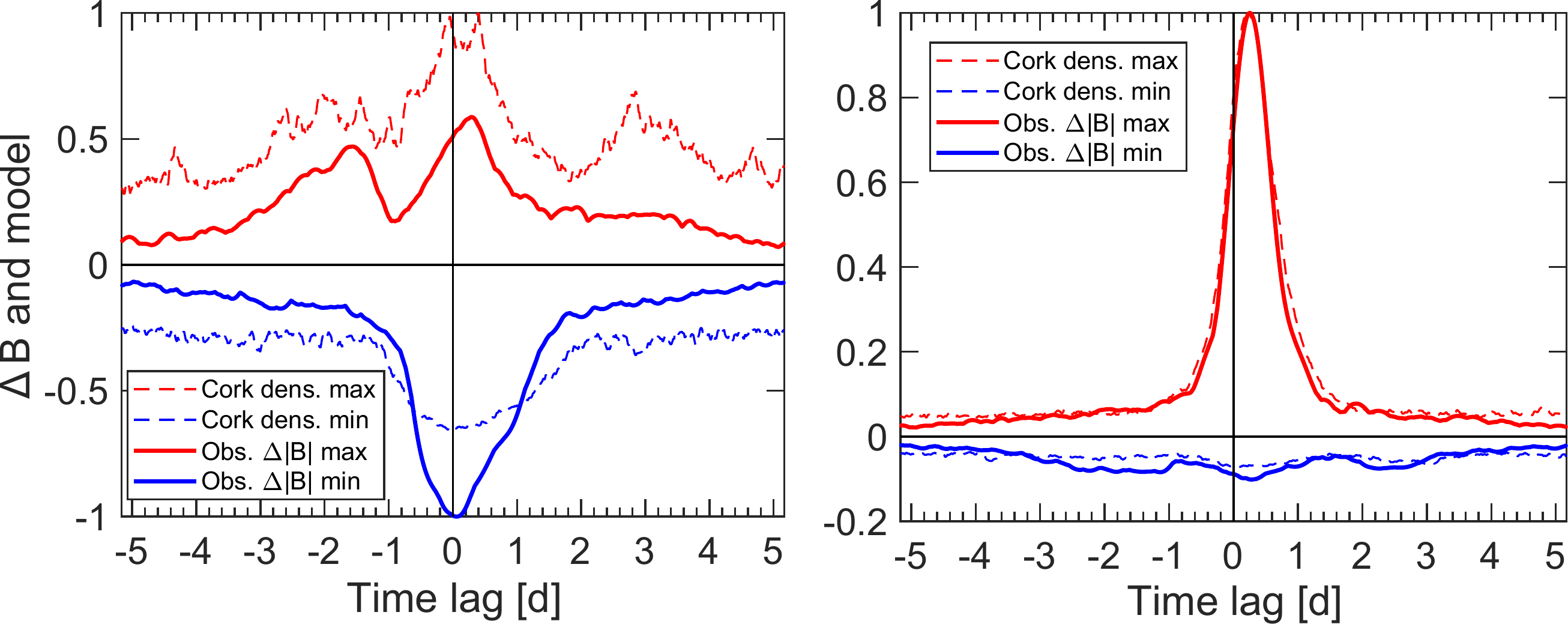}
\caption{Maximum and minimum magnetic field (observations) and cork density (simulation) over each average supergranule map as a function of time lag. \textit{Left:} Outflow. \textit{Right:} Inflow. The values are shown after subtraction of mean and division by the maximum absolute value to facilitate the comparison. The cork density is taken from a simulation run with $\eta = 250~$km$^2$~s$^{-1}$ and $t_\text{life} = 16$~h (run \#5).}
\label{fig_cork-timelag}
    \end{figure*}

Despite the limitations of the simulation, the evolution of the magnetic field distribution (Fig.~\ref{fig_B}) is well reproduced by the corks (see Fig.~\ref{fig_cork}).\footnote{Movies of the magnetic field evolution and cork simulation can be found on-line. See Appendix~\ref{sect_movies} for additional information.} In particular, the time lag of the strongest cork concentration in the average supergranular inflow matches the time lag of the strongest magnetic field concentration (see Fig.~\ref{fig_cork-timelag}). The spatial peak size is reduced though (FWHM is about 5~Mm compared to 7~Mm).

%______________________________________________ cork simulation and B: evolution of maxima/minima

To analyze how sensitive the time lag is to the cork simulation parameters, we performed a parameter study. We varied the turbulent diffusivity and the cork lifetime, running nine simulations for various combinations of these parameters (see Table~\ref{table_timelag} in the Appendix). The original simulation run with $\eta = 250~\text{km}^2~\text{s}^{-1}$ and $t_\text{life} = 16$~h serves as our reference (run \#5 in the table).

For each simulation run and the observed magnetic field, we measured the time of highest cork density and strongest magnetic field with respect to the time of strongest horizontal inflow. To this end, we constructed the time series of maximum cork density and magnetic field values in each frame (as shown by the red curves in Fig.~\ref{fig_cork-timelag}) and used two different fitting methods to obtain the time lag from these time series: a parabolic fit using only the three temporal pixels around the maximum, and a Lorentzian fit using the entire time series. The resulting time lags for the different parameter combinations are shown in Table~\ref{table_timelag}. All simulation runs produce positive time lags in the range between 3.6 and 7.7~hours, with the majority lying within one hour of the observed time lag for the magnetic field. Smaller differences are likely without strong meaning, judging from the differences between the two fitting methods, which  are typically 0.5$-$1~hours. However, there are (weak) trends of decreasing time lags for increasing diffusivity, and increasing time lags for increasing lifetimes. This results in an underestimation of the time lag for run \#7 and an overestimation for runs \#2 and \#3.

\begin{figure*}[h]
\sidecaption
\includegraphics[height=5.0cm]{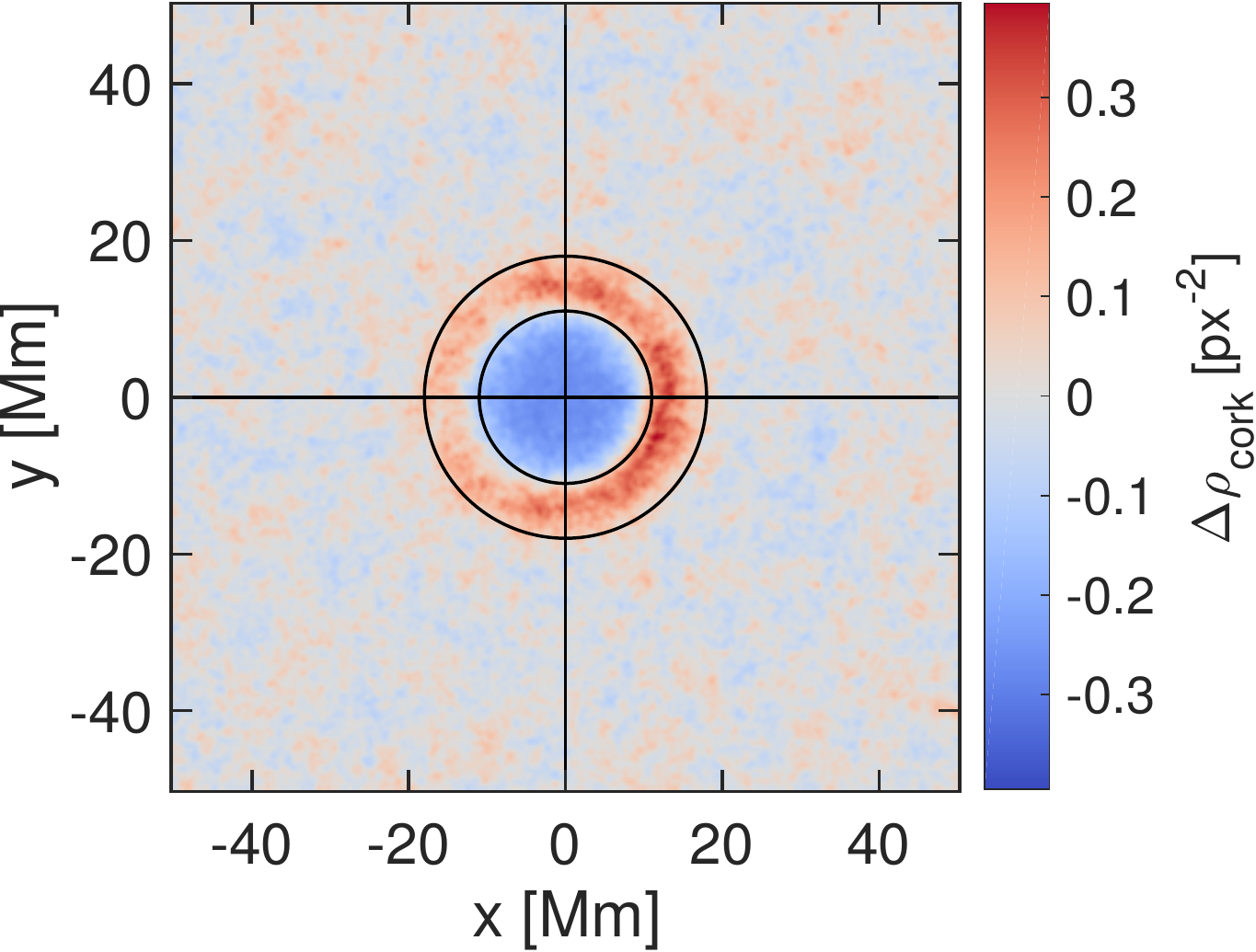}
\includegraphics[height=5.0cm]{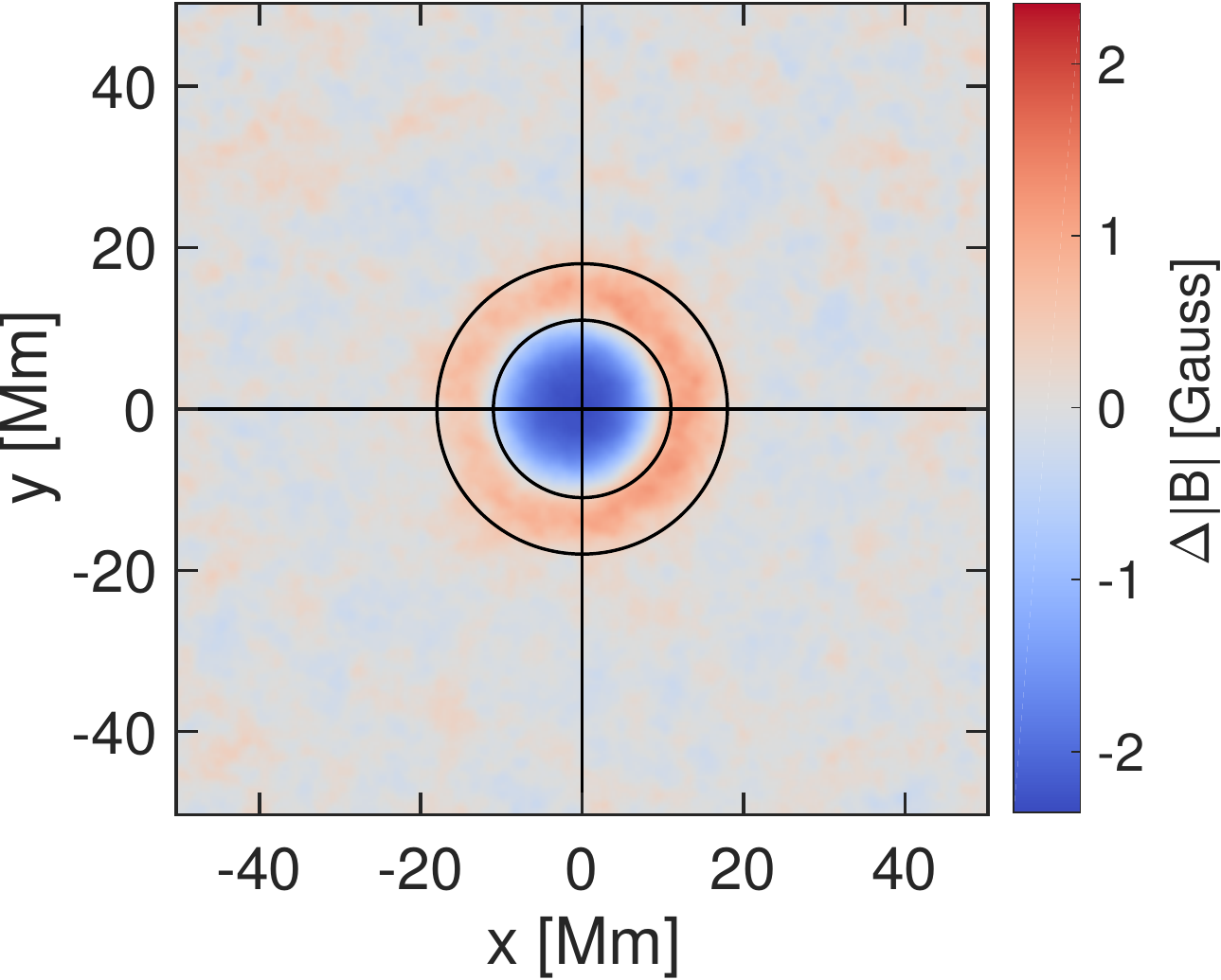}
\caption{Simulated cork density (\textit{left}) and observed magnetic field (\textit{right}) for the average supergranular outflow at reference time ($\Delta t = 0$), after subtraction of mean. The cork density is taken from a simulation run with $\eta = 250~\text{km}^2~\text{s}^{-1}$ and $t_\text{life} = 16$~h (run \#5). The circles have radii 11~Mm and 18~Mm with respect to the origin.}
\label{fig_cork-dt0_outflow}
    \end{figure*}

For the average supergranular outflow, we can qualitatively reproduce the clearing of the magnetic field in the outflow center, however with a shallower minimum compared to the observed field (see Fig.~\ref{fig_cork-dt0_outflow}, for simulation run \#5). Around the outflow, a ring of increased cork density forms, corresponding to the magnetic network field. This ring of higher cork density stands out more clearly than for the observed magnetic field.

%______________________________________________ cork simulation and B: stills at dt=0 for outflow (east-west anisotropy)

%______________________________________________ cork simulation and B: still at dt=0 for outflow (east-west anisotropy) for different eta and t_life values
   \begin{figure*}[h]
\centering
\includegraphics[height=4.3cm]{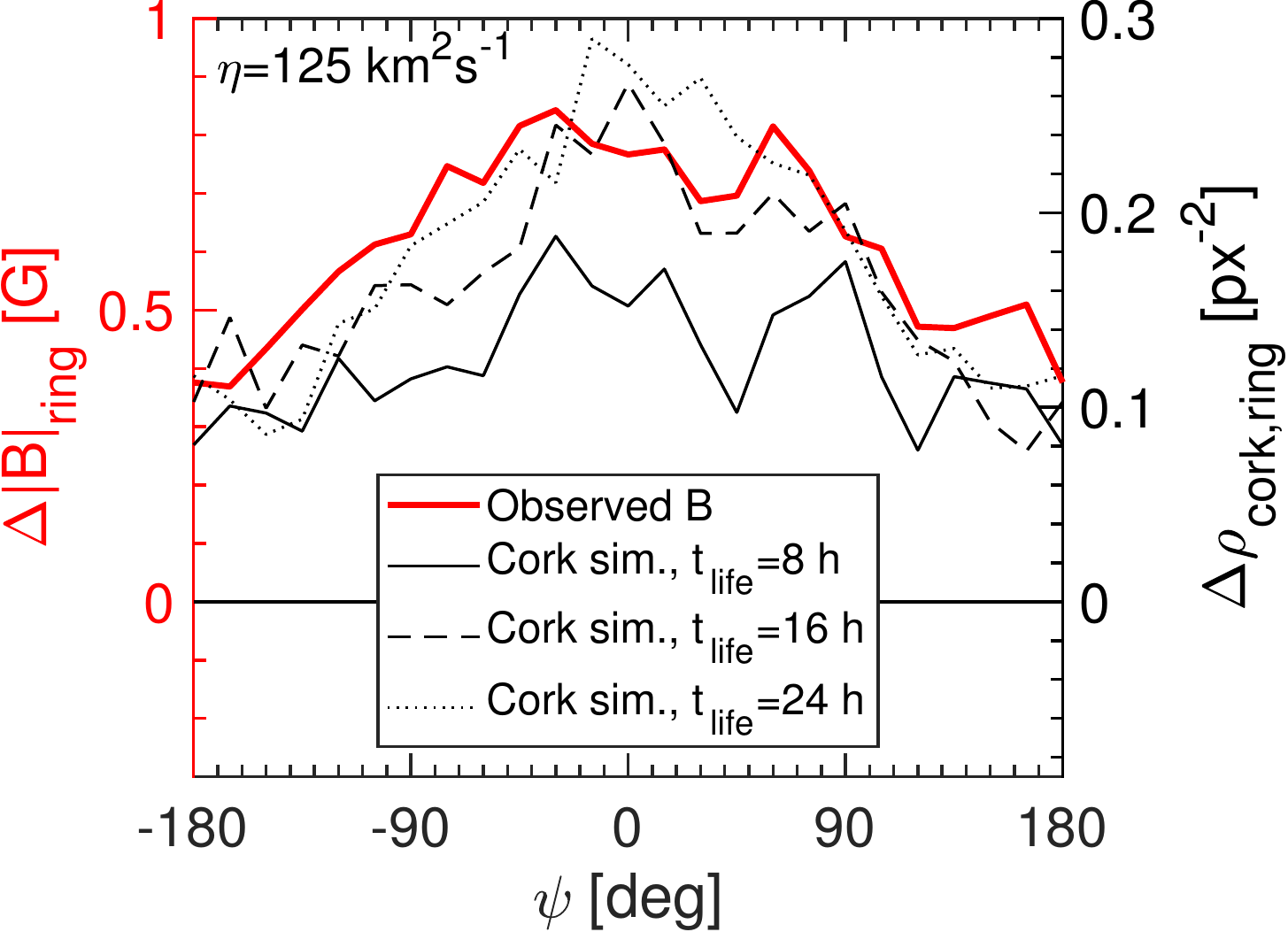}
\includegraphics[height=4.3cm]{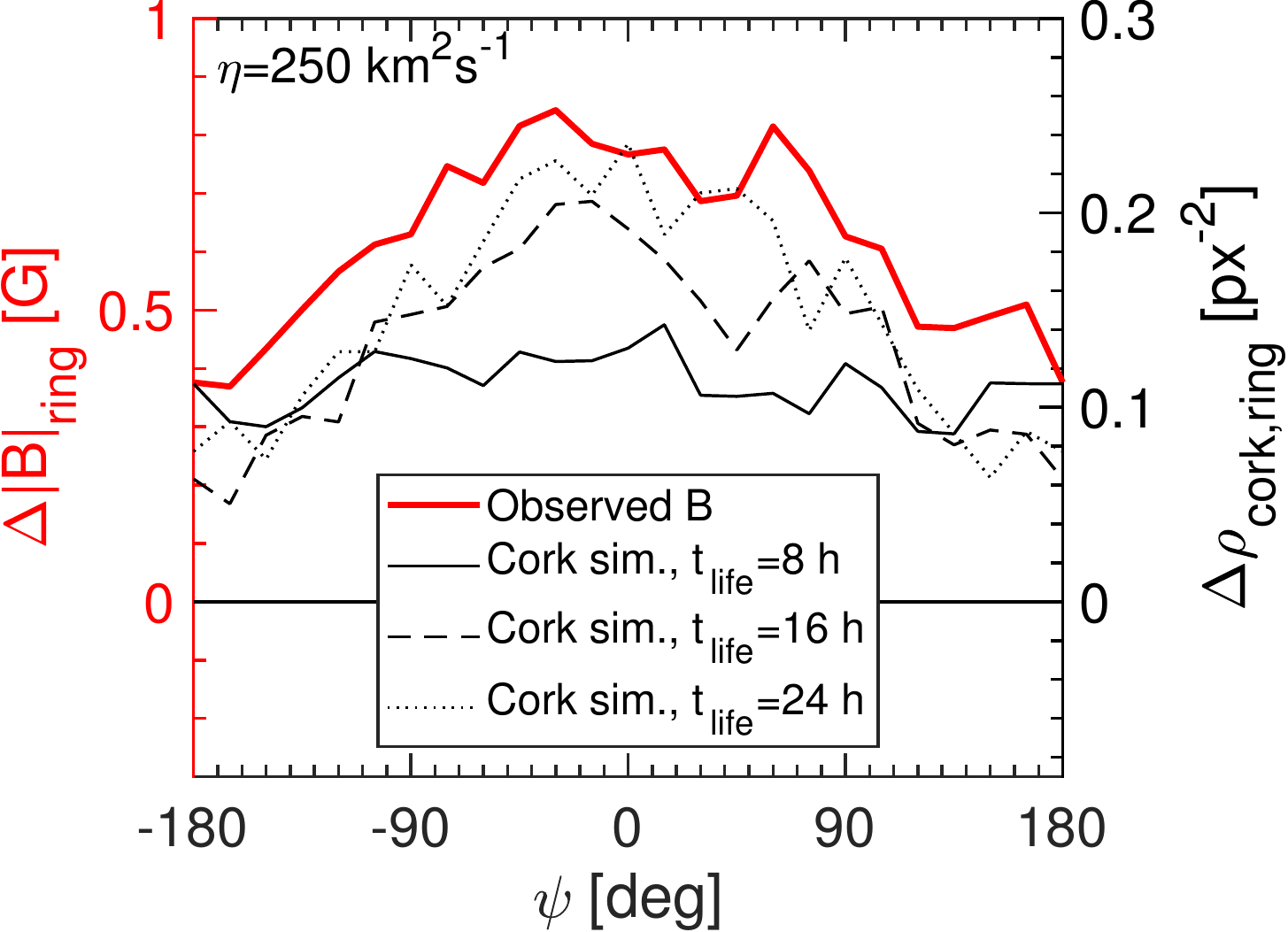}
\includegraphics[height=4.3cm]{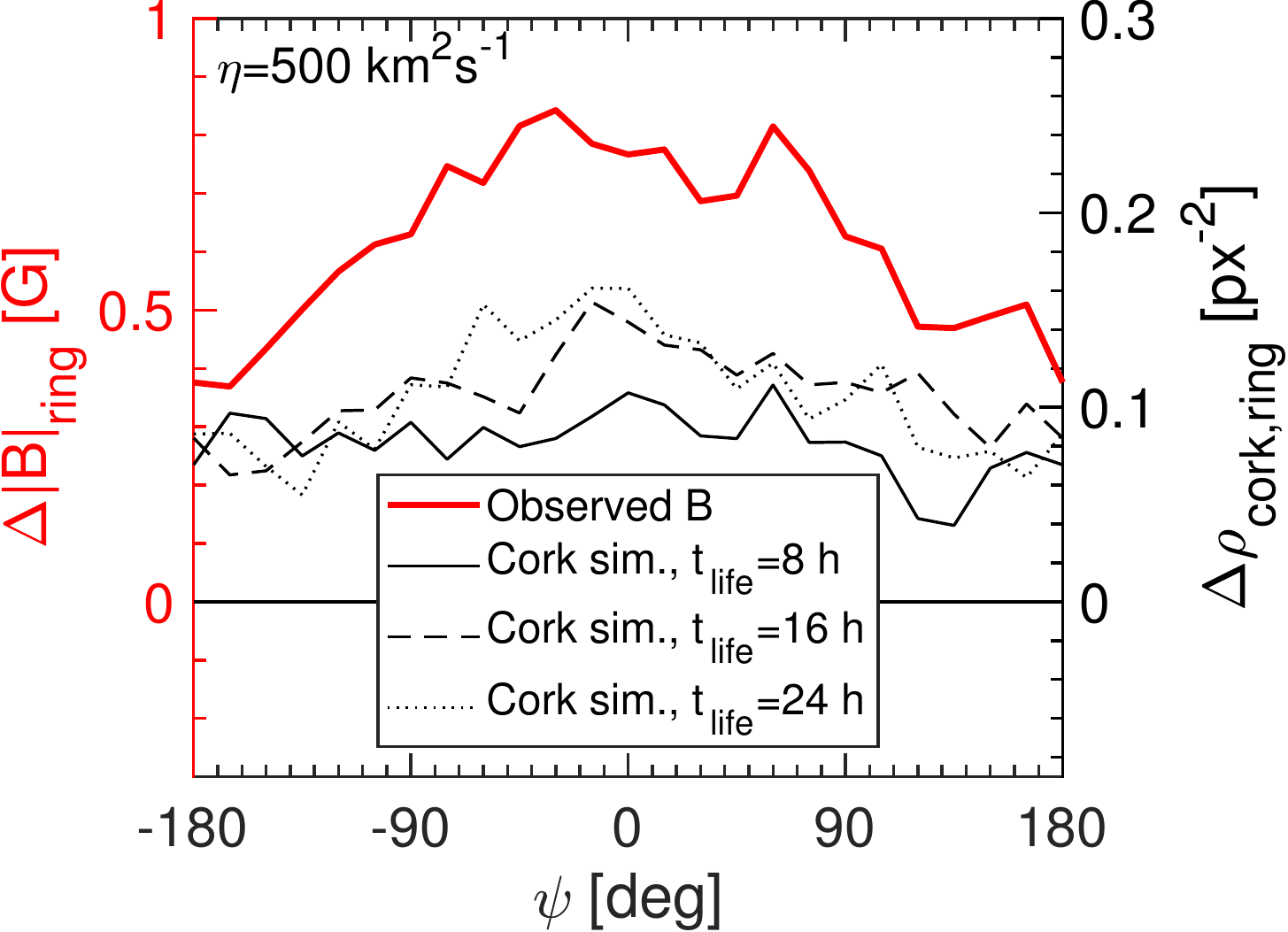}
\caption{Simulated cork density in a ring surrounding the outflow center ($ 11~\text{Mm} < r < 18~\text{Mm}$, as shown in Fig.~\ref{fig_cork-dt0_outflow}) as a function of azimuthal angle, $\psi$. Results are shown for nine separate simulation runs with different combinations of $\eta$ and $t_\text{life}$ parameter values (see Table~\ref{table_timelag}). The angle $\psi=0$ points west. All results are at reference time ($\Delta t = 0$), after subtraction of the mean over the image. We show the observed magnetic field in all three panels for reference.}
\label{fig_cork-dt0_outflow_params}
    \end{figure*}
    
The observed network field is stronger on the west side of the average supergranule than on the east side, as we detected in our previous study \citep{langfellner_2015a} and as was later also found by \citet{roudier_2016}. This observation is successfully reproduced by the cork simulation (run \#5).
To study the dependence of this anisotropy on the simulation parameters, we analyzed the azimuthal distribution of the network field by averaging the cork density in a ring (see circles in Fig.~\ref{fig_cork-dt0_outflow}) around the outflow center over azimuthal bins of $15\degr$.
The results for all simulation runs are shown in Fig.~\ref{fig_cork-dt0_outflow_params}. The anisotropy is similar to the observed magnetic field for all runs with $t_\text{life} = 16~$h and 24~h. For shorter lifetimes ($t_\text{life} = 8$~h), the anisotropy is much less pronounced, if present at all. In those cases, the corks apparently vanish before the supergranular flows had enough time to aggregate the corks on the western side of the supergranules.
This is consistent with the observation by \citet{gosic_2014} that the net contribution of the internetwork field to the network field is positive (amplification stronger than cancellation) and the most important contribution factor, with a flux replacement time of 18$-$24~h.
Also, a higher than observed diffusivity ($\eta = 500~$km$^2$~s$^{-1}$) results in a weaker anisotropy. 
Overall, these results support our previous speculation that the supergranular oscillatory flows acting on the magnetic field are able to produce the network anisotropy.

%%%%%%%%%%%%%%%%%%%%%%%%%%%%%%%%%%%%%%%%%%%%%%%%%%%%%%%%%%%%%%%

\section{Summary and discussion}

\subsection{Supergranular wave power and evolution}
We have confirmed the supergranular power spectrum of the divergence signal measured by \citet{gizon_2003} and \citet{schou_2003} at the solar equator and extended it to higher wavenumbers, with a good agreement between TD and LCT of granules. As was found in the previous studies, the wave power is strongest in the prograde (westward) direction. Lorentzian fits to the power spectrum yield at $kR_\odot > 120$ a strong deviation from the empirical square-root-of-$k$ dispersion relation in form of a flat profile with an oscillation frequency of ${\sim}2~\mu$Hz. For $kR_\odot > 270$, the line width is too broad to extract a meaningful dispersion relation.

For the average supergranule in real space, we detected a complex evolution pattern with an oscillatory component, both with TD and LCT. Average supergranular outflows and inflows turn into each other with a mean period of 6$-$7~days, viewed at a tracking rate corresponding to the supergranulation pattern rotation rate (Carrington rate plus $40$~m~s$^{-1}$). This means that new supergranules are preferentially born in inflow regions west of existing supergranules. 

Indeed, the supergranular evolution pattern is scale-dependent; restricting the selection of supergranules to narrow size ranges in the averaging process reveals an increase of the oscillation period and lifetime for larger cells. These trends are consistent with the parameter values extracted from the Lorentzian fits of the supergranular power spectra. We can thus relate the signature of supergranular waves in Fourier space to the evolution pattern of the average supergranule in real space.

Changing the tracking rate to a lower value (Carrington rate minus~20~m~s$^{-1}$) introduces a westward drift of the evolution pattern corresponding to the tracking speed difference ($60$~m~s$^{-1}$), but does not alter the oscillation pattern. In Fourier space, this drift corresponds to a Doppler frequency shift $ku_x$.

\subsection{Are supergranules deep?}
As we showed in Fig.~\ref{fig_power-fit-dispersion}, the background flow $u_x$ shows a slight dependence on spatial scale. A potential explanation for this might be that supergranules of different sizes are rooted at different depths and are sensitive to the change of the rotation rate in the near-surface shear layer.

To get a rough idea if this might be the case, we relate $u_x(k)$ to measurements of the solar rotation rate from global-mode helioseismology.
Let us assume that supergranules have a depth $z$ that depends linearly on their horizontal extent $L$, with a scaling factor $\alpha$, i.e. $L = \alpha z$. With $L = 2\pi/k$ and $z=R_\odot - r_\text{base}$, where $r_\text{base}$ is the radial distance from the solar center to the base of the supergranule, this gives a relationship between $kR_\odot$ and $r_\text{base}$ that depends on $\alpha$:
\begin{equation}
kR_\odot = \frac{2\pi}{\alpha \left(1 - r_\text{base}/R_\odot \right)} .  \label{eq_kR}
\end{equation}

%______________________________________________ v_rot from SG power spectra fits and 2D RLS inversion (Larson and Schou, 2018)
   \begin{figure*}[h]
\sidecaption
%\centering
\includegraphics[width=0.7\hsize]{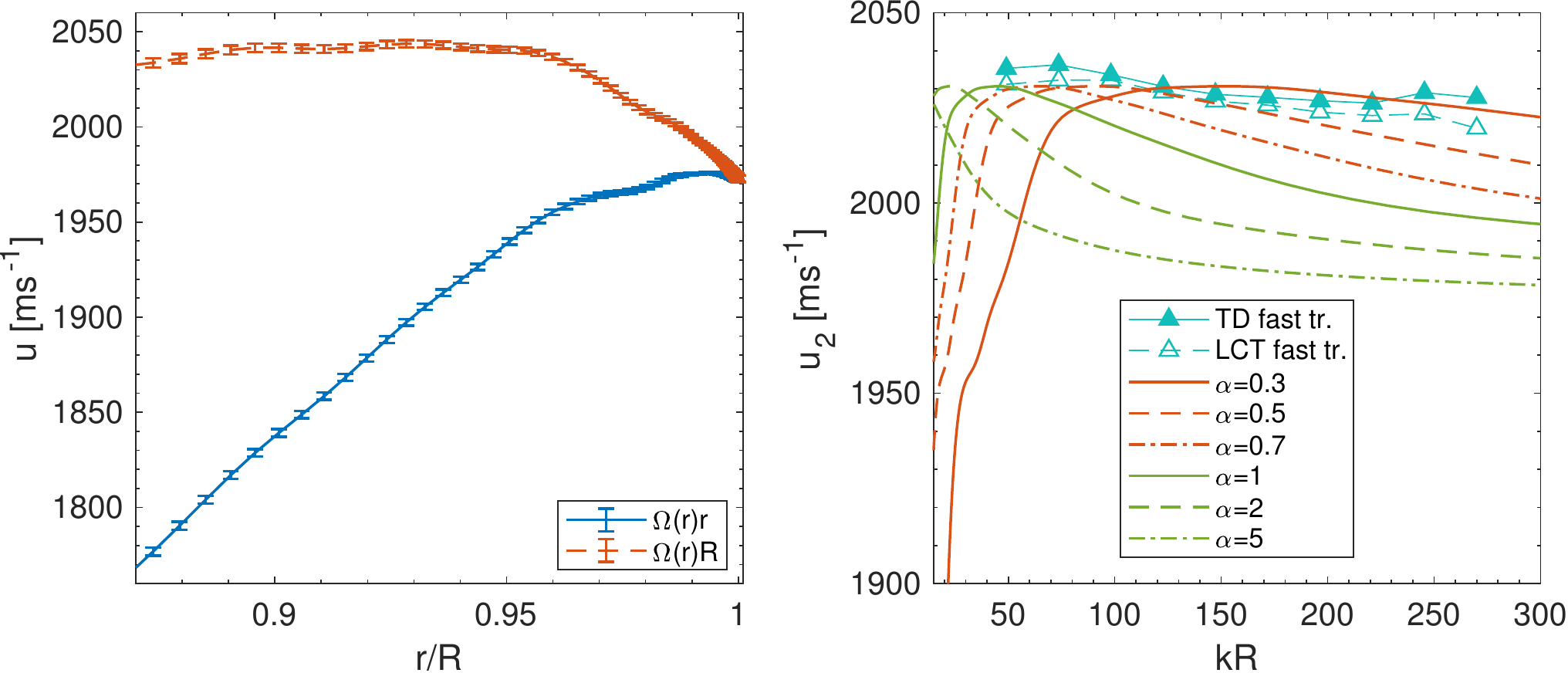}
\caption{\textit{Left:} Linear rotational velocity $\Omega(r)r$ and rotation rate $\Omega(r)R_\odot$ vs.~radial distance from solar center at solar equator obtained from a global-mode inversion \citep{larson_2018}. \textit{Right:} Vertical average of rotation rate $\Omega(r)R_\odot$ between base of supergranule and solar surface vs.~$kR_\odot$, for various $\alpha$ values (see text for definitions of $u_2$ and $\alpha$). Also shown is $u_x$ plus tracking speed (2034~m~s$^{-1}$) from supergranular power spectrum fits (TD and LCT fast tracking). Note that the $y$-axis range is different in both panels.}
\label{fig_vrot}
    \end{figure*}

We further speculate that the velocity $u_x(k)$ that is "felt" by a supergranule is some weighted average of the rotation rate $\Omega(r)$ in the range of $r$ between the top and the base of the supergranule. Let us consider two scenarios for the weights. 

In the first scenario, we suppose that supergranules feel an average of the \textit{linear velocity} $\Omega(r)r$ between $r_\text{base}$ and the solar surface ($R_\odot$):
\begin{equation}
u_1(r_\text{base}) = \frac{\int_{r_\text{base}}^{R_\odot} \Omega(r)r \text{d}r}{R_\odot - r_\text{base}} .
\end{equation}
In a second scenario, motivated by the idea that the motion of supergranules at the surface reflects the angular velocity at their rooting depths, we suppose that supergranules feel the average of the \textit{angular velocity} $\Omega(r)$. This gives a higher weight to greater depths compared to the first scenario, as $\Omega(r)$ is multiplied by the solar radius $R_\odot$ at all depths instead of the local radius $r \leq R_\odot$:
\begin{equation}
u_2(r_\text{base}) = \frac{\int_{r_\text{base}}^{R_\odot} \Omega(r)R_\odot \text{d}r}{R_\odot - r_\text{base}} .
\end{equation}

The left panel of Fig.~\ref{fig_vrot} shows the linear rotational velocity $\Omega(r)r$ and the rotation rate $\Omega(r)$ (multiplied by $R_\odot$ for ease of comparison), where $\Omega(r)$ is the equatorial rotation rate from a 2D global-mode RLS inversion of HMI Dopplergrams \citep[360-day average in 2010--11,][]{larson_2018}.
As the rotational velocity never exceeds 1980~m~s$^{-1}$ for any depth, the weighted average $u_1$ is lower than the supergranular velocity (see right panel) by at least 50~m~s$^{-1}$, irrespective of $\alpha$.

On the other hand, the quantity $\Omega(r)R_\odot$ reaches 2040~m~s$^{-1}$. With Eq.~\ref{eq_kR}, we can express the weighted average $u_2$ as a function of horizontal scale $kR_\odot$. It is shown in Fig.~\ref{fig_vrot} for different values of $\alpha$, together with $u_x$ from the supergranular power fits where the fast tracking speed (2034~m~s$^{-1}$) was added.
While only considering the angular velocity $\Omega(r_\text{base})R_\odot$ at the base of the supergranule would yield projected velocity values that are 10~m~s$^{-1}$ too high, the average $u_2$ between $r_\text{base}$ and $R_\odot$ matches the supergranular velocity profile quite closely for $\alpha \sim 0.4$. In our simple picture, this corresponds to vertically elongated supergranules, e.g. $L = 30~$Mm would yield a depth $z \sim 75~$Mm, which is deeper than the near-surface shear layer.

While this match might be pure coincidence, it motivates further studies of the supergranular wave properties, e.g. using depth inversions, to explore the possibility of deep-rooted supergranules, as originally suggested by \citet{parker_1973}. We note that other recent works typically reported shallow supergranules \citep[e.g.,][]{rieutord_2010a, duvall_2014}.

\subsection{Passive magnetic field}
We found that the magnetic field undergoes similar oscillations as the horizontal divergent flows on supergranular scales, but with a time lag of about 6~hours. The magnetic field evolution, including the time lag, can be well reproduced with a simple simulation, where magnetic elements are modeled as corks. In this simulation, the corks are advected by the average supergranular flows, possess a finite lifetime and are affected by smaller-scale motions through the turbulent diffusivity.
The cork simulation is also able to reproduce the previously observed east-west anisotropy of the network field \citep{langfellner_2015a,roudier_2016} surrounding the average supergranule.

The occurrence of the 6-hour time lag both in the magnetic field observations and the cork simulation is indicative of the passive nature of the magnetic field in supergranules and confirms the findings of \citet{simon_1989a}, \citet{orozco_2012} and \citet{roudier_2016}. In our measurements and simulations, there is no sign of magnetic fields shaping the supergranular flows as proposed by \citet{crouch_2007} or network field arising from random walks \citep{thibault_2012}. Instead, the higher cork densities in our simulations, which correspond to the network field, arise solely due to the advection by the supergranular flows.

However, an influence of the magnetic fields on the (quiet-Sun) supergranular flows in certain cases cannot be completely excluded. For example, \citet{sangeetha_2016} found that the vorticity in supergranular inflows is reduced if the magnetic field is strong. According to \citet{kobel_2012}, in the strongest parts of the magnetic network the convection can be inhibited, causing magnetic elements to appear darker than in regions with a weaker magnetic field. It is unclear though how far the influence of the magnetic field goes.

Attempts to relate the magnetic field strength to supergranulation parameters are difficult; e.g., \citet{chatterjee_2017} reported that the supergranule size increases in active regions with stronger magnetic field, whereas a strong quiet-Sun network field lets supergranules shrink. \citet{meunier_2007b}, on the other hand, found that a strong \textit{network} field is associated with \textit{bigger} cells and a strong \textit{internetwork} field is responsible for \textit{smaller} cells. Thus, the results depend on the exact definition and measurement place of the magnetic field strength and are difficult to compare and interpret.

%%%%%%%%%%%%%%%%%%%%%%%%%%%%%%%%%%%%%%%%%%%%%%%%%%%%%%%%%%%%%%%%%%%%%%%%%%%%%%%%%
\begin{acknowledgements}
The HMI data used are courtesy of NASA/SDO and the HMI science team.
The data were processed at the German Data Center for SDO (GDC-SDO), funded by the German Aerospace Center (DLR).
We thank J.~Schou, R.~Cameron and T.~Duvall, Jr. for the useful discussions.
We are grateful to R.~Burston and H.~Schunker for providing help with the data processing, especially the tracking and mapping. We used the workflow management system Pegasus funded by The National Science Foundation under OCI SI2-SSI program grant \#1148515 and the OCI SDCI program grant \#0722019 as well as the distributed computing software HTCondor.
\end{acknowledgements}

\bibliographystyle{aa}
\bibliography{literature}

%%%%%%%%%%%%%%%%%%%%%%%%%%%%%%%%%%%%%%%%%%%%%%%%%%%%%%%%%%%%%%%%%%%%%%%%%%%%%%%
%%%%%%%%%%%%%%%%%%%%%%%%%%%%%%%%%%%%%%%%%%%%%%%%%%%%%%%%%%%%%%%%%%%%%%%%%%%%%%%

\appendix

\section{Additional plots and tables}

\subsection{Azimuthal dependence of power}
\begin{figure*}[h]
\sidecaption
\parbox{0.7\hsize}{
\hspace{0.2\hsize} \framebox{\Large{TD}}  \hspace{0.4\hsize}  \framebox{\Large{LCT}} \\
\includegraphics[page=1,width=0.499\hsize]{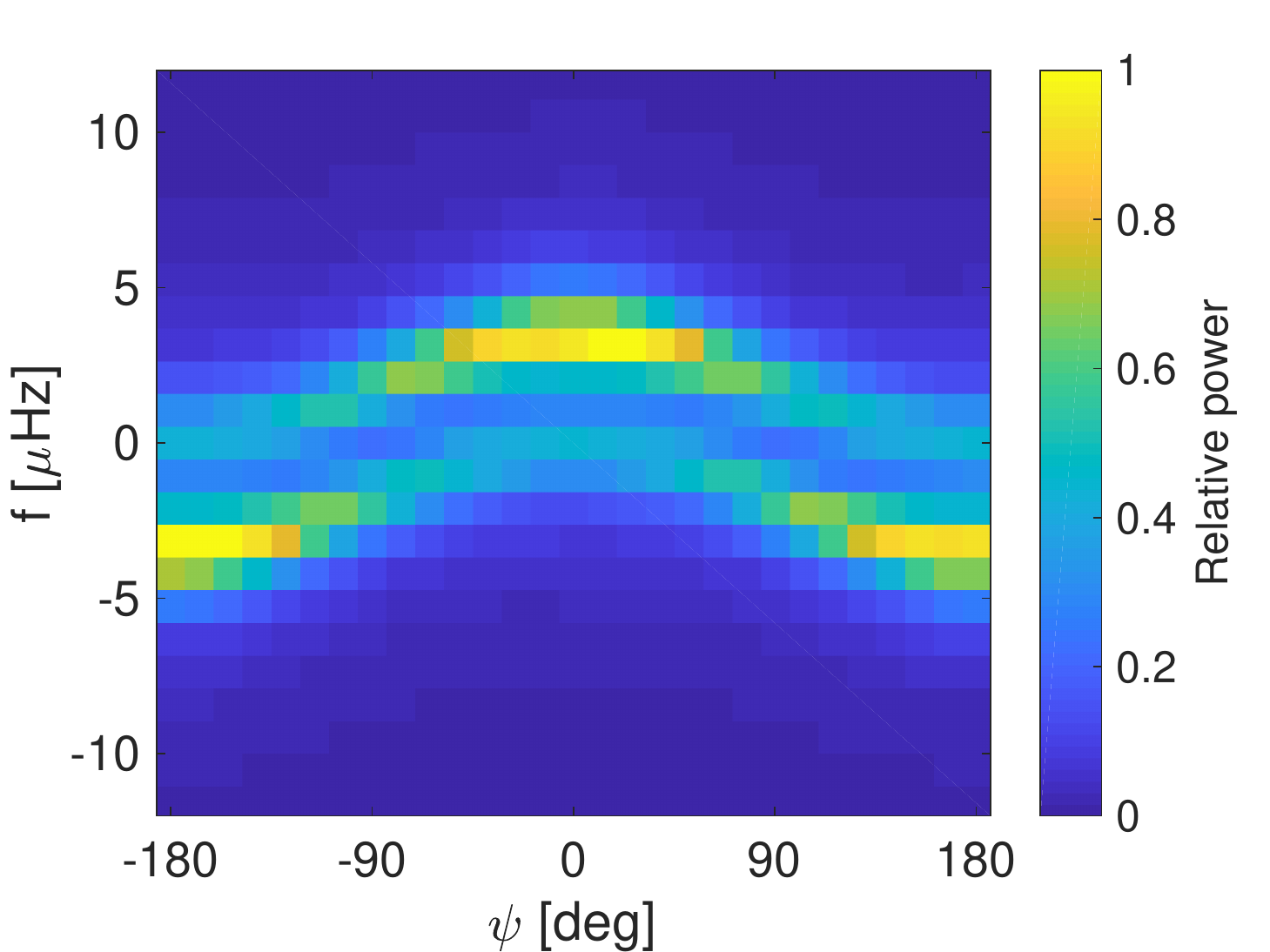}
\includegraphics[page=2,width=0.499\hsize]{figA1}
\includegraphics[page=3,width=0.499\hsize]{figA1}
\includegraphics[page=4,width=0.499\hsize]{figA1}
}
\caption{Supergranular power spectrum at the solar equator: Azimuthal dependence at $kR_\odot \approx 123$ for the \citeauthor{snodgrass_1984} tracking rate using TD (f modes) and LCT. \textit{Top row:} Observed power. \textit{Bottom row:} Lorentzian fit. Frequencies beyond $\pm7~\mu$Hz are not taken into account.}
\label{fig_sg-power_azimuth_snodgrass}
    \end{figure*}

\begin{figure*}[h]
\sidecaption
\parbox{0.7\hsize}{
\hspace{0.2\hsize} \framebox{\Large{TD}}  \hspace{0.4\hsize}  \framebox{\Large{LCT}} \\
\includegraphics[page=1,width=0.499\hsize]{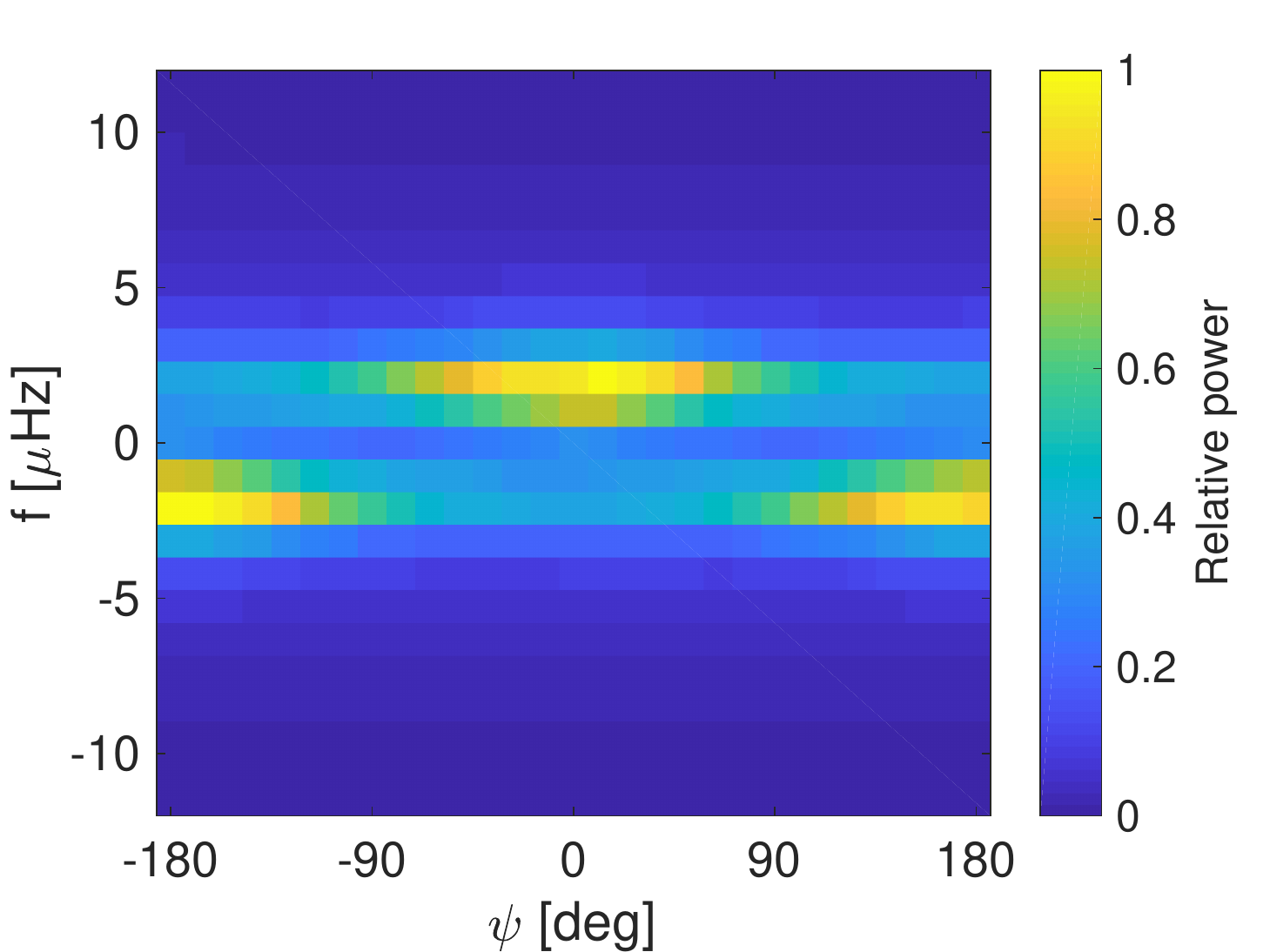}
\includegraphics[page=2,width=0.499\hsize]{figA2}
\includegraphics[page=3,width=0.499\hsize]{figA2}
\includegraphics[page=4,width=0.499\hsize]{figA2}
}
\caption{As Fig.~\ref{fig_sg-power_azimuth_snodgrass}, but for the fast tracking rate.}
\label{fig_sg-power_azimuth_fast}
    \end{figure*}

\begin{figure*}[h]
\sidecaption
\parbox{0.7\hsize}{
\hspace{0.2\hsize} \framebox{\Large{TD}}  \hspace{0.4\hsize}  \framebox{\Large{LCT}} \\
\includegraphics[page=1,width=0.499\hsize]{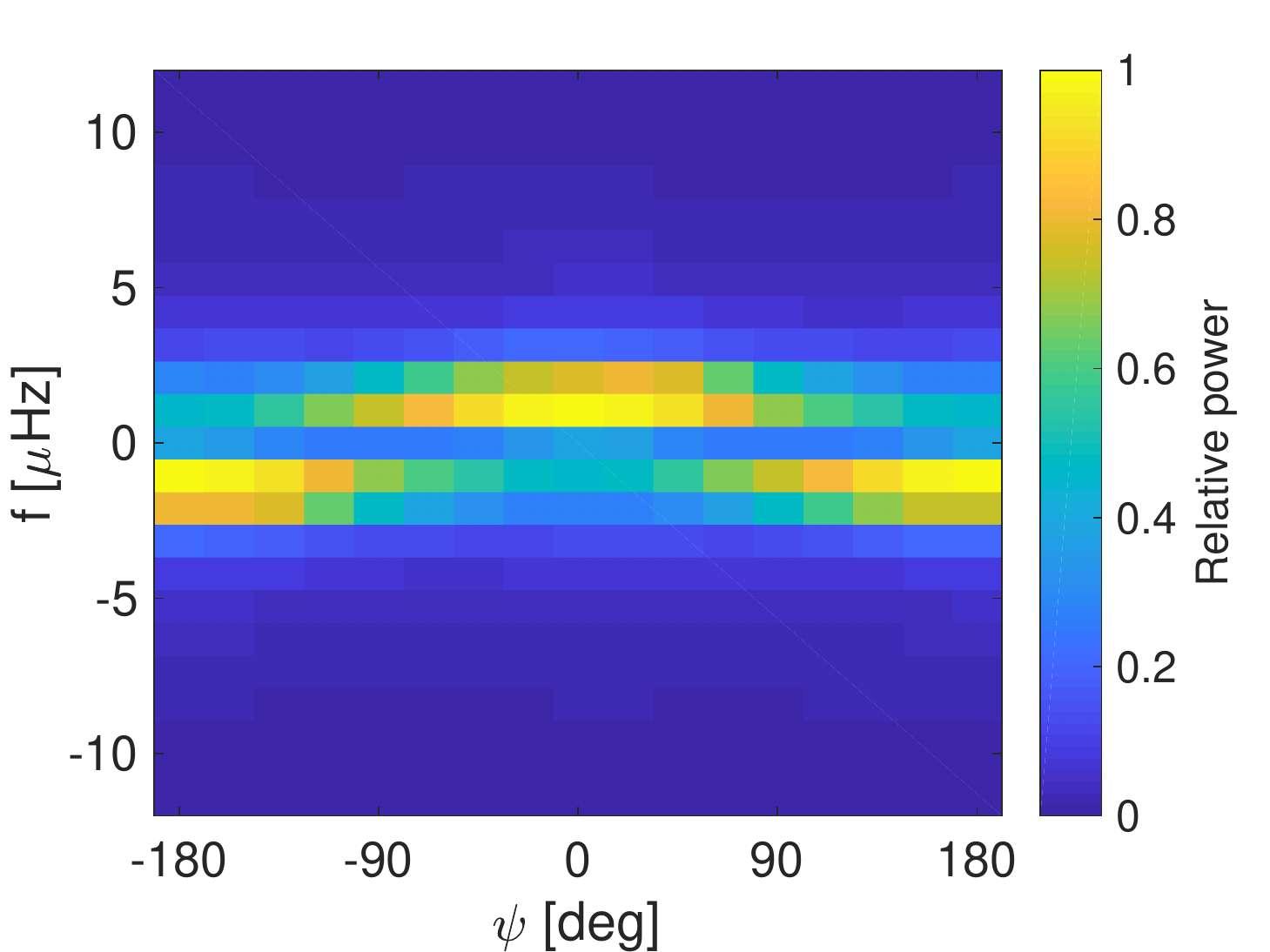}
\includegraphics[page=2,width=0.499\hsize]{figA3}
\includegraphics[page=3,width=0.499\hsize]{figA3}
\includegraphics[page=4,width=0.499\hsize]{figA3}
}
\caption{As Fig.~\ref{fig_sg-power_azimuth_fast}, but for $kR_\odot = 74$.}
\label{fig_sg-power_azimuth_fast_kR74}
    \end{figure*}

\begin{figure*}[h]
\sidecaption
\parbox{0.7\hsize}{
\hspace{0.2\hsize} \framebox{\Large{TD}}  \hspace{0.4\hsize}  \framebox{\Large{LCT}} \\
\includegraphics[page=1,width=0.499\hsize]{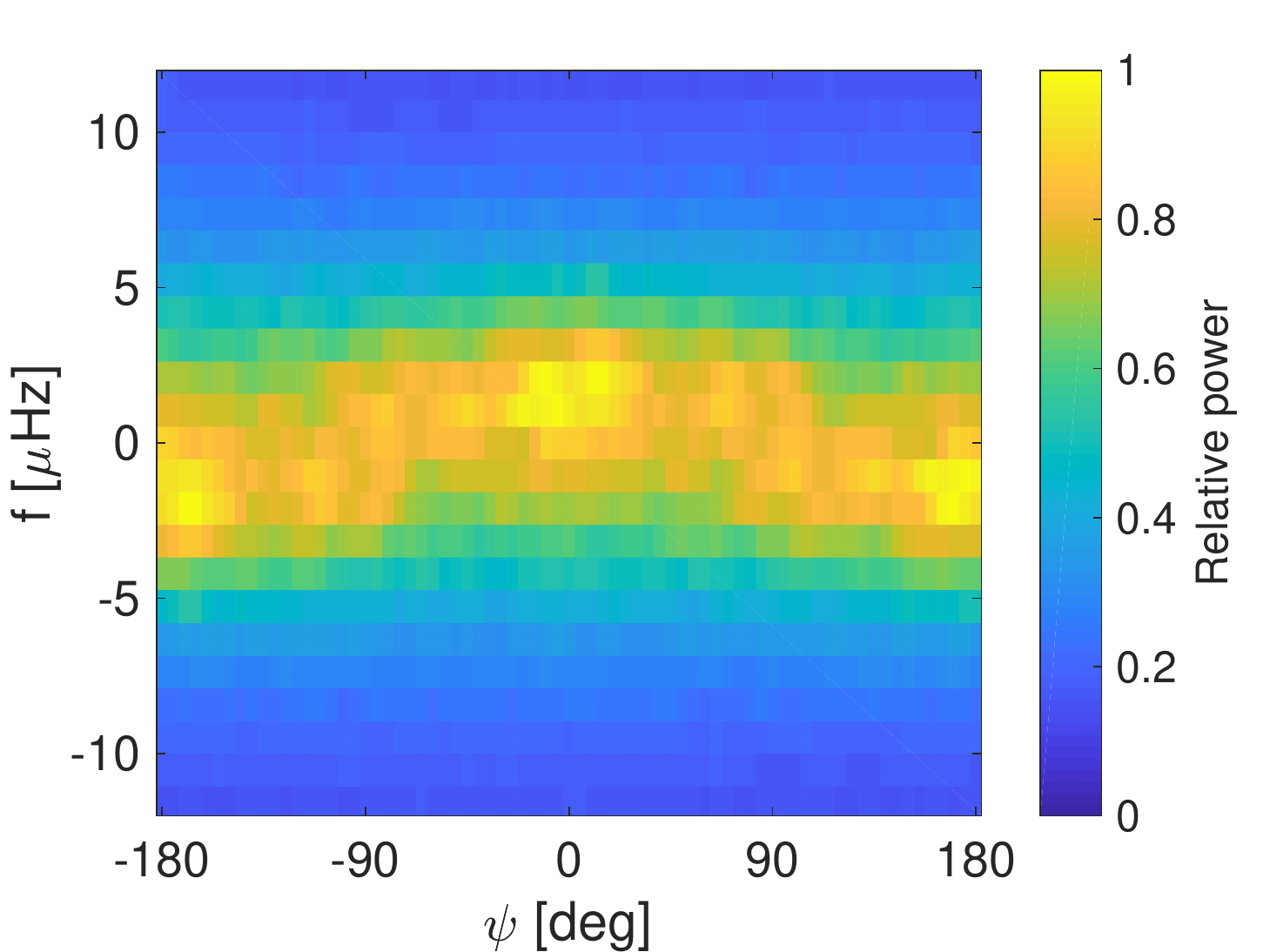}
\includegraphics[page=2,width=0.499\hsize]{figA4}
\includegraphics[page=3,width=0.499\hsize]{figA4}
\includegraphics[page=4,width=0.499\hsize]{figA4}
}
\caption{As Fig.~\ref{fig_sg-power_azimuth_fast}, but for $kR_\odot = 270$.}
\label{fig_sg-power_azimuth_fast_kR270}
    \end{figure*}

This Appendix consists of Figs.~\ref{fig_sg-power_azimuth_snodgrass} through \ref{fig_sg-power_azimuth_fast_kR270}.

%______________________________________________ SG power spectra: azimuthal dependence (Snodgrass tracking) kR=123

%______________________________________________ SG power spectra: azimuthal dependence (fast tracking) kR=123

%______________________________________________ SG power spectra: azimuthal dependence (fast tracking) kR=74

%______________________________________________ SG power spectra: azimuthal dependence (fast tracking) kR=270

\newpage

%%%%%%%%%%%%%%%%%%%%%%%
\subsection{Fit parameters}
This Appendix consists of Table~\ref{table_fit-parameters}.

%______________________________________________ Snodgrass tracking
\begin{table}[h]
\centering
\caption{Parameters from Lorentzian power fit.}
\label{table_fit-parameters}
\begin{tabular}{c c c c c c}
\hline\hline
$kR_\odot$ & f$_0$ & HWHM & t$_\text{life}$ & $u_x$ & $u_y$ \\
 & [$\mu$Hz] & [$\mu$Hz] & [d] & [m s$^{-1}$] & [m s$^{-1}$] \\ \hline
  49&   1.14&   0.78&   2.38&  66.9&   1.2 \\
  74&   1.46&   0.80&   2.32&  61.9&   1.9 \\
  98&   1.71&   0.84&   2.20&  57.5&  -0.1 \\
 123&   1.89&   0.99&   1.85&  54.8&   1.1 \\
 147&   2.00&   1.24&   1.49&  51.8&   0.3 \\
 172&   2.04&   1.64&   1.12&  51.0&  -1.2 \\
 196&   2.03&   2.10&   0.88&  48.8&  -0.5 \\
 221&   2.04&   2.68&   0.69&  47.0&   0.3 \\ \hline
\end{tabular}
\tablefoot{The values are for TD with \citet{snodgrass_1984} tracking rate. Errors for each scale $kR_\odot$ can be estimated from the spread of curves in Figs.~\ref{fig_power-fit-dispersion} and \ref{fig_fit-parameters}.}
\end{table}

\newpage

%%%%%%%%%%%%%%%%%%%%%%%
\subsection{Evolution: dependence on tracking rate and spatial scale}
\begin{figure*}[h]
\centering
\includegraphics[width=\hsize]{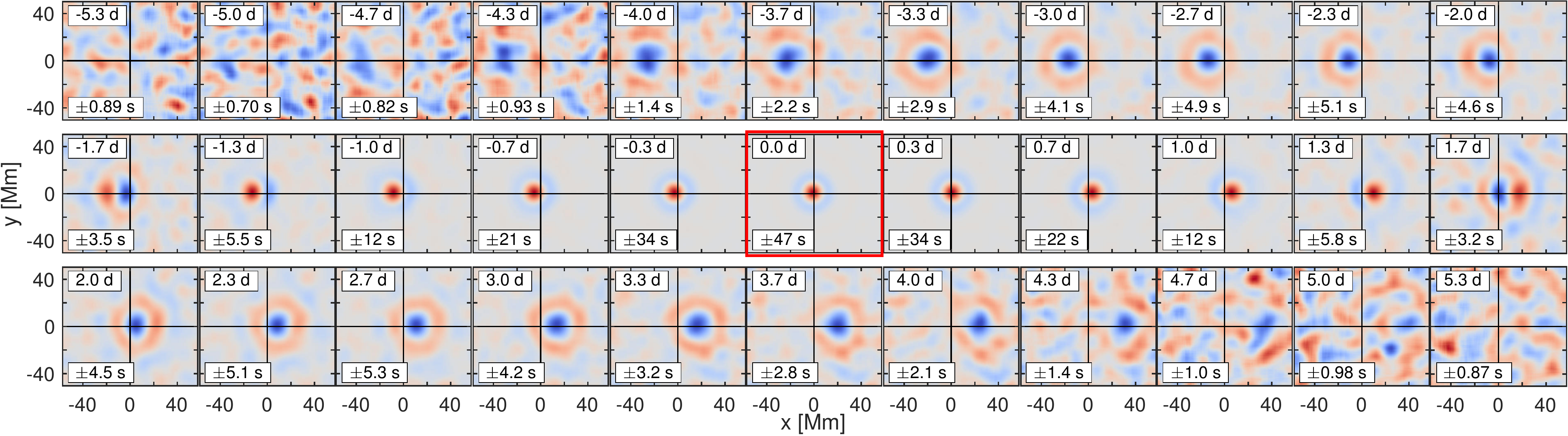} \\
\caption{Evolution of f-mode-filtered inward$-$outward point-to-annulus travel times for the average supergranular outflow: \citet{snodgrass_1984} tracking rate (cf.~Fig.~\ref{fig_tauoi}: tracking at supergranulation pattern rotation rate, which is faster by about $60$~m~s$^{-1}$).}
\label{fig_tauoi_tracking}
    \end{figure*}

This Appendix consists of Fig.~\ref{fig_tauoi_tracking}.

%______________________________________________ tau^oi evolution for avg. SG outflow: slower tracking rate

%%%%%%%%%%%%%%%%%%%%%%%
\subsection{Cork simulation details}  \label{sect_corksim-details}
The cork simulations that we ran to model the magnetic field on supergranular scales (see Sect.~\ref{sect_corksim}) span 11~days in time (as long as the observation datasets) with a time step $\Delta t_\text{sim} = 15~\text{min}$ on a domain of size $150\times150$~Mm$^2$. The simulations were initialized with $N_\text{init} \approx 1.8\times10^5$ corks at random positions of uniform distribution, corresponding to one cork per square pixel on average. This number effectively sets the noise level; the more corks, the lower the noise. At each time step, the simulations were advanced by applying the following actions:
\begin{enumerate}
\item Advection of existing corks by horizontal supergranular background flow. We used the time-dependent $\vec{u} = (u_x,u_y)$ measured with LCT for the average supergranule that was also used to compute $\text{div}_\text{h} \vec{u}$ shown in Fig.~\ref{fig_LCTdiv}.
\item Random walk of existing corks. The average distance was determined by the turbulent diffusivity $\eta = 250~$km$^2$~s$^{-1}$ \citep{simon_1997} for the standard run. For other runs, we also used 125 and $500~$km$^2$~s$^{-1}$ (see Table~\ref{table_timelag}).
\item Corks that leave the simulation domain are removed.
\item The remaining corks have a probability $p_\text{dec} = 1- \exp(-\Delta t_\text{sim}/t_\text{life})$ to be removed from the simulation (exponential decay). The cork lifetime $t_\text{life}$ was chosen to be equal to the flux replacement time $t_\text{repl} \sim 16~\text{h}$ \citep{hagenaar_2003} for the standard run. For other runs, we also used 8 and 24~h.
\item To compensate for the removed corks and to simulate the emergence of new flux, we spawn $N_\text{spawn} = p_\text{dec} N_\text{init}$ new corks at random positions with uniform probability.
\end{enumerate}

%_____________________________________________ cork simulation: dependence of time lag peak location on parameters (eta, t_life)
   \begin{table}
     \caption{Time lag of simulated cork density and observed magnetic field.} 
\label{table_timelag}      
\centering                          
\begin{tabular}{c|c c|c c}        
\hline\hline                 
Simulation & $\eta$ & $t_\text{life}$ & \multicolumn{2}{c}{Time lag [h]} \\
run & [km$^2$ s$^{-1}$] & [h] & Parabolic fit & Lorentzian fit \\
\hline
\#1 & 125 & 8 & 6.5 & 5.8 \\
\#2 & 125 & 16 & 7.7 & 7.1 \\
\#3 & 125 & 24 & 7.5 & 7.7 \\
\#4 & 250 & 8 & 6.5 & 5.3 \\ \hline
\#5 (ref.) & 250 & 16 & 5.4 & 6.2 \\ \hline
\#6 & 250 & 24 & 6.5 & 6.6 \\
\#7 & 500 & 8 & 4.1 & 3.6 \\
\#8 & 500 & 16 & 5.9 & 5.1 \\
\#9 & 500 & 24 & 3.7 & 6.0 \\
\hline
Magnetic & & & & \\
field & 500 & 24 & 6.1 & 6.3 \\
\hline 
\end{tabular}
   \end{table}

%%%%%%%%%%%%%%%%%%%%%%%
\newpage
\section{Evolution movies}  \label{sect_movies}
\begin{figure*}[h]
\sidecaption
\includegraphics[width=0.7\hsize]{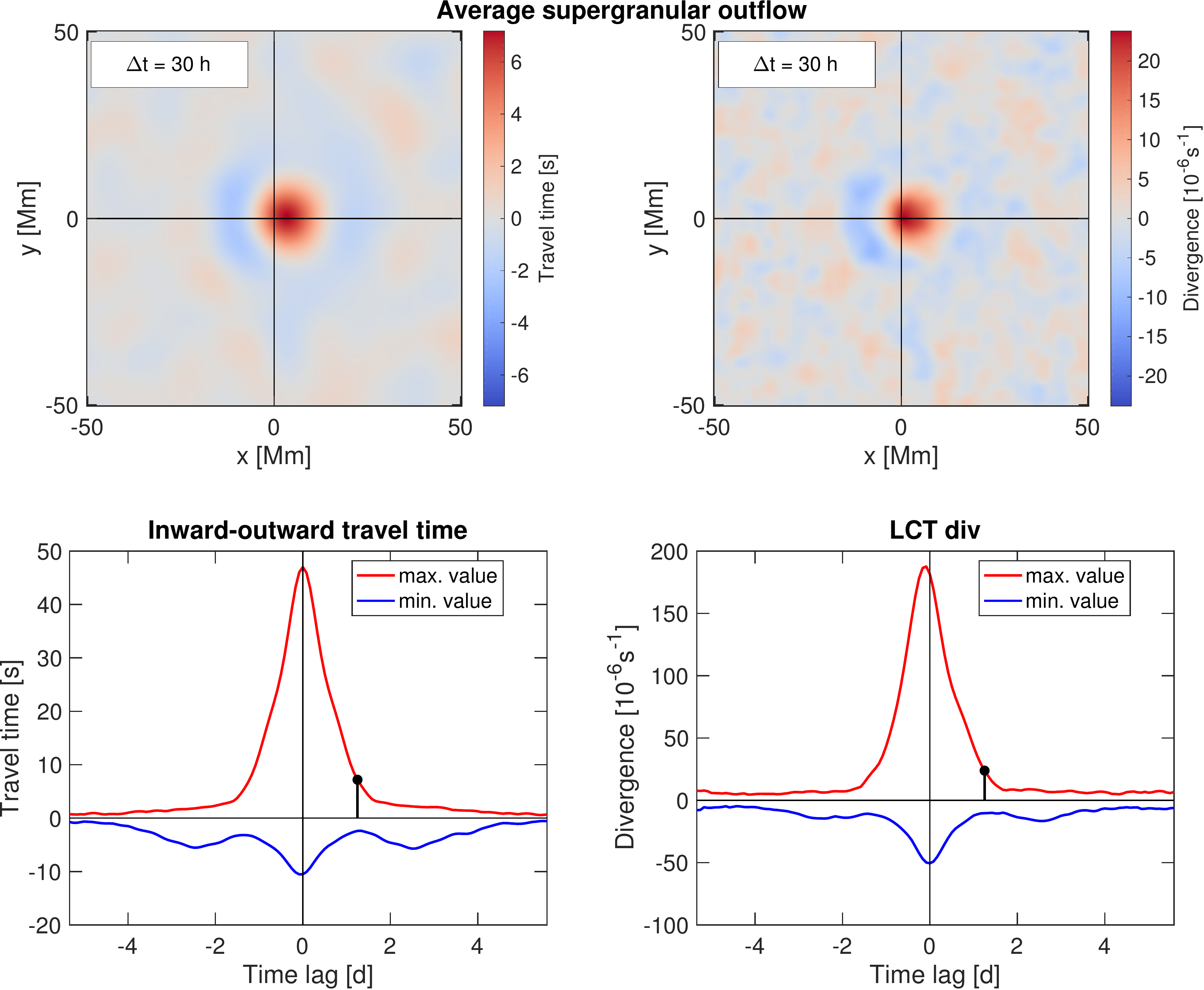}
\caption{Still from supergranular temporal evolution movie at time lag $\Delta t = 30~$h. \textit{Top row:} TD (inward$-$outward travel times) and LCT $\text{div}_h$ for the average supergranular outflow, as shown in Figs.~\ref{fig_tauoi} and \ref{fig_LCTdiv}. Frames at time lags that are not multiples of 8~hours were calculated using temporal Fourier interpolation. \textit{Bottom row:} Maximum and minimum travel time and LCT $\text{div}_h$ over each average supergranule map as a function of time lag. The highest absolute value at the current time lag is marked with the black circle. This value is reflected by the limits of the colorbar in the top row panels. The full temporal evolution for both, the inflow and the outflow, can be found in the two movies attached to this figure.}
\label{fig_TD-LCT-movie}
    \end{figure*}

This paper is accompanied by four movies that show the temporal evolution of the average supergranule. Fig.~\ref{fig_TD-LCT-movie} shows a still from the evolution movie of the average supergranular outflow, which juxtaposes TD and LCT (see caption for details). The corresponding movie for the average supergranular inflow (no still shown) is composed analogously.

\begin{figure*}[h]
\sidecaption
\includegraphics[width=0.7\hsize]{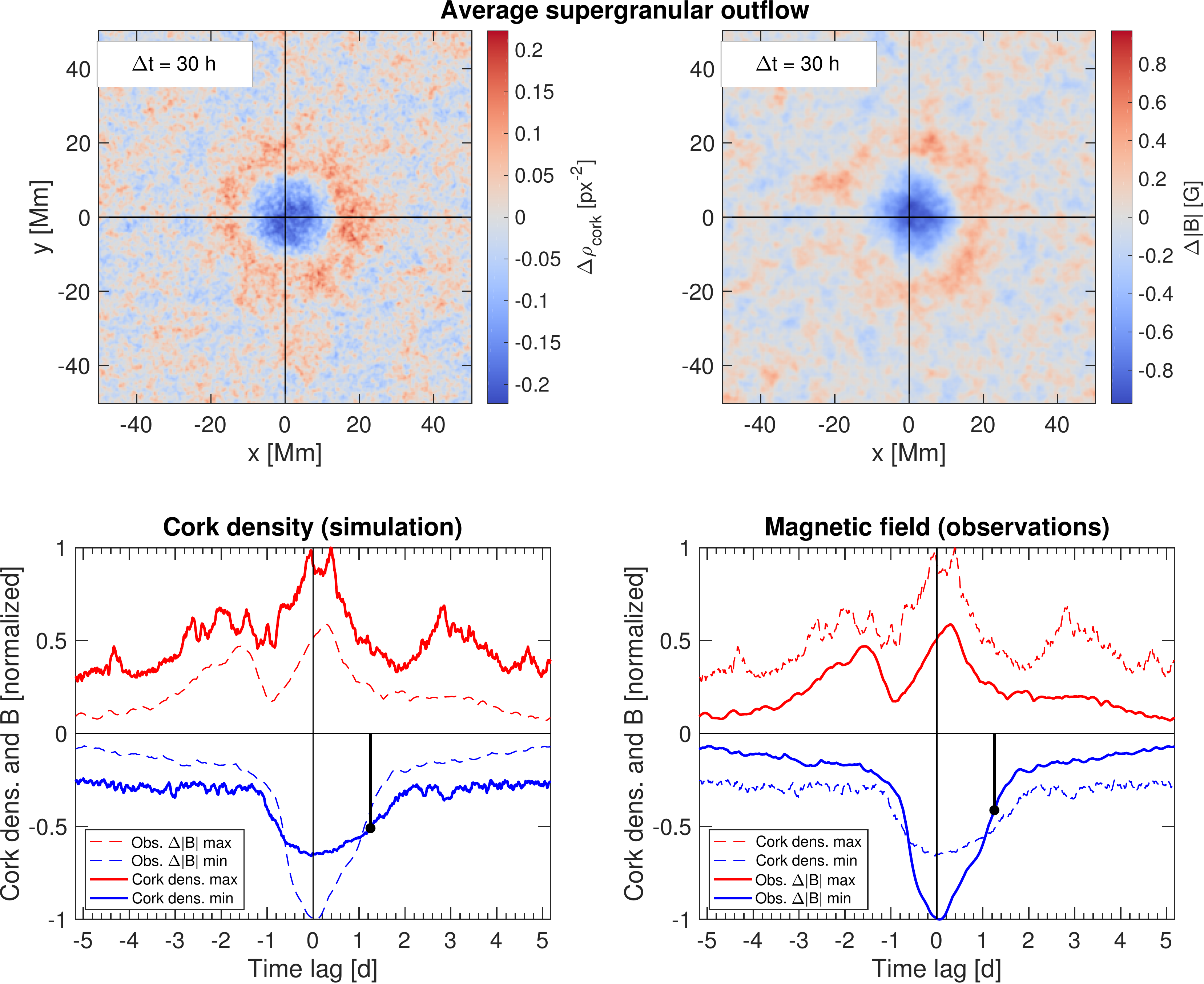}
\caption{Still from supergranular evolution movie. \textit{Top row:} Simulated cork density (run \#5) and observed line-of-sight magnetic field for the average supergranular outflow, as shown in Figs.~\ref{fig_cork} and \ref{fig_B}. \textit{Bottom row:} Maximum and minimum cork density and magnetic field strength as in Fig.~\ref{fig_cork-timelag}. In addition, the highest absolute value at the current time is marked with the black circle. The full temporal evolution for both, the inflow and the outflow, can be found in the two movies attached to this figure.}
\label{fig_cork-B-movie}
    \end{figure*}

Fig.~\ref{fig_cork-B-movie} shows a still from the movie that directly compares the observed evolution of the line-of-sight magnetic field and the cork simulation (run \#5) for the average supergranular outflow. In this case, too, the corresponding movie for the average supergranular inflow has the same structure, and thus no still is shown.

%______________________________________________ TD oi and LCT div evolution for avg. SG outflow
    
%______________________________________________ Cork simulation and observed B for avg. SG outflow

\section{ArXiv appendix}
This Appendix includes figures that are not part of the A\&A publication.

%______________________________________________ tau^oi evolution for avg. SG outflow (kR=52, 124, 222)
   \begin{figure*}[h]
\centering
\includegraphics[width=\hsize]{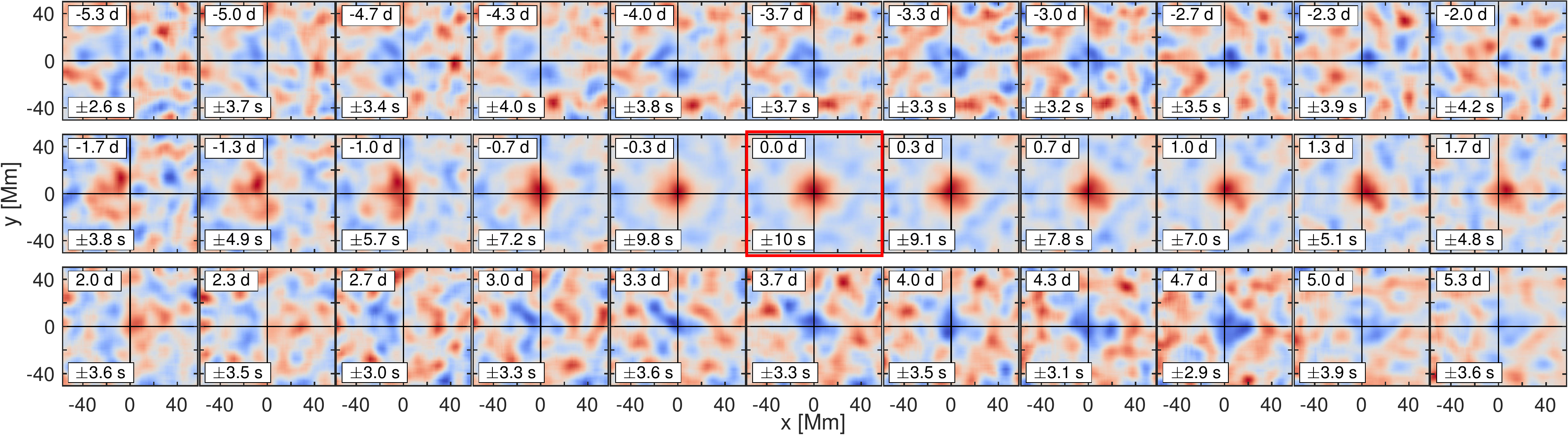} \\
\includegraphics[width=\hsize]{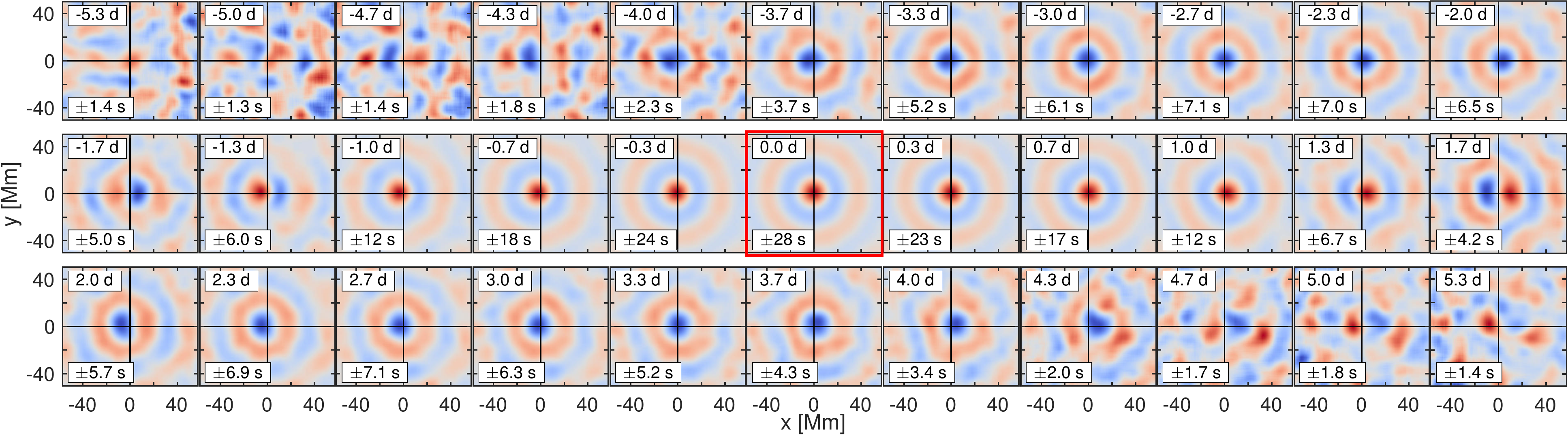} \\
\includegraphics[width=\hsize]{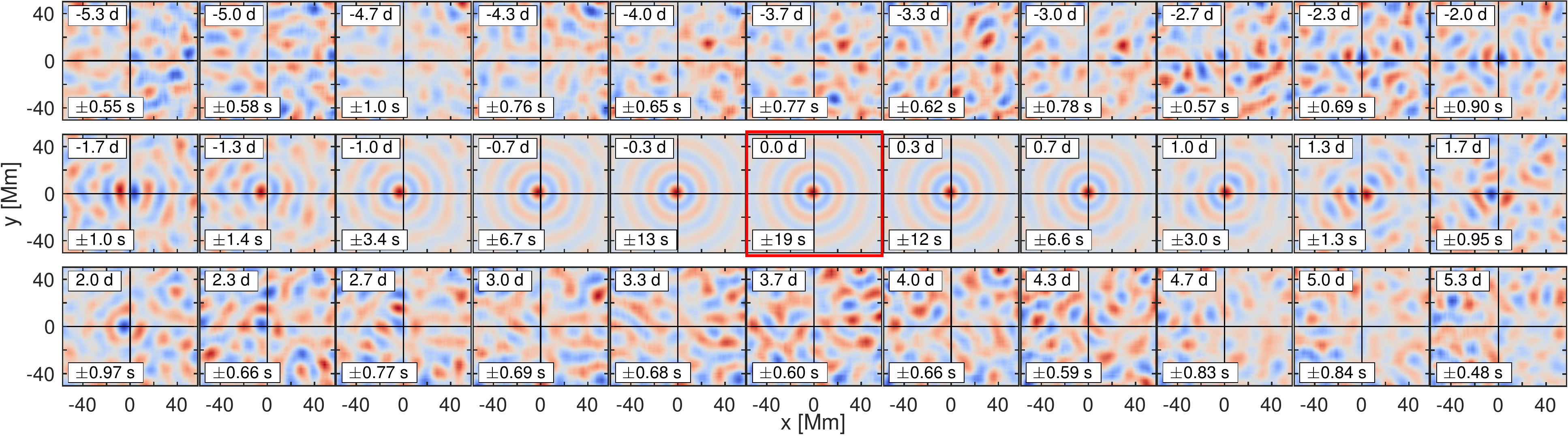}
\caption{Evolution of f-mode-filtered inward$-$outward travel times for the average convective outflow (cf.~Fig.~\ref{fig_tauoi}, top panels) at three different scales. \textit{From top to bottom:} $kR_\odot = 52$ (${\sim}$80~Mm), $kR_\odot = 124$ (${\sim}$35~Mm), $kR_\odot = 222$ (${\sim}$20~Mm). The spatial band-pass filters are Gaussian in $kR_\odot$ with $\sigma \approx 12$.}
\label{fig_tauoi_kRscales_outflow}
    \end{figure*}

%______________________________________________ tau^oi evolution for avg. SG inflow (kR=52, 124, 222)
   \begin{figure*}[h]
\centering
\includegraphics[width=\hsize]{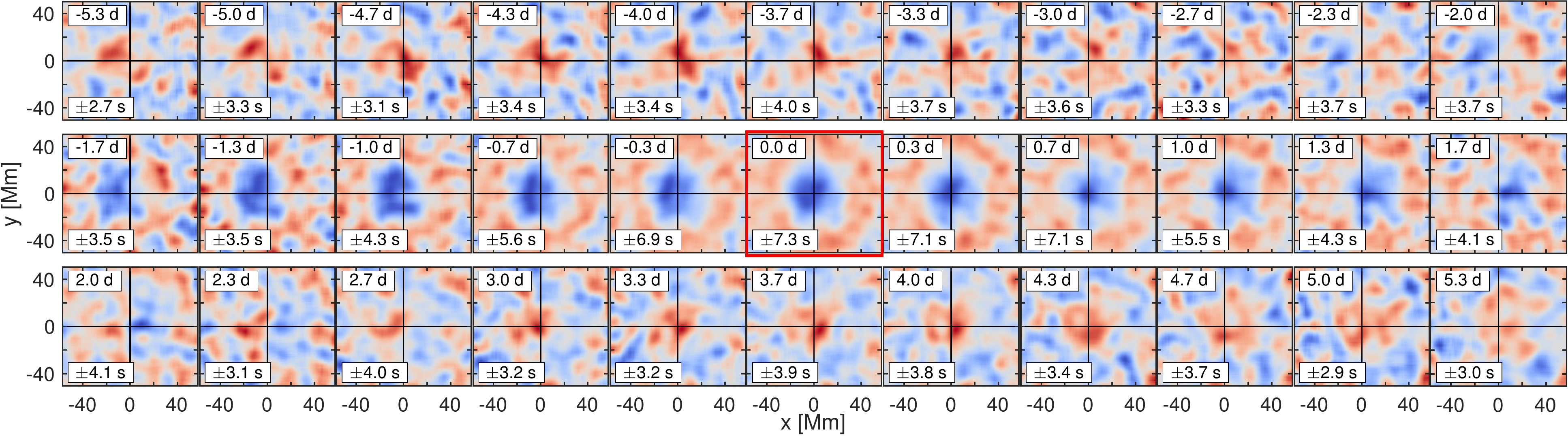} \\
\includegraphics[width=\hsize]{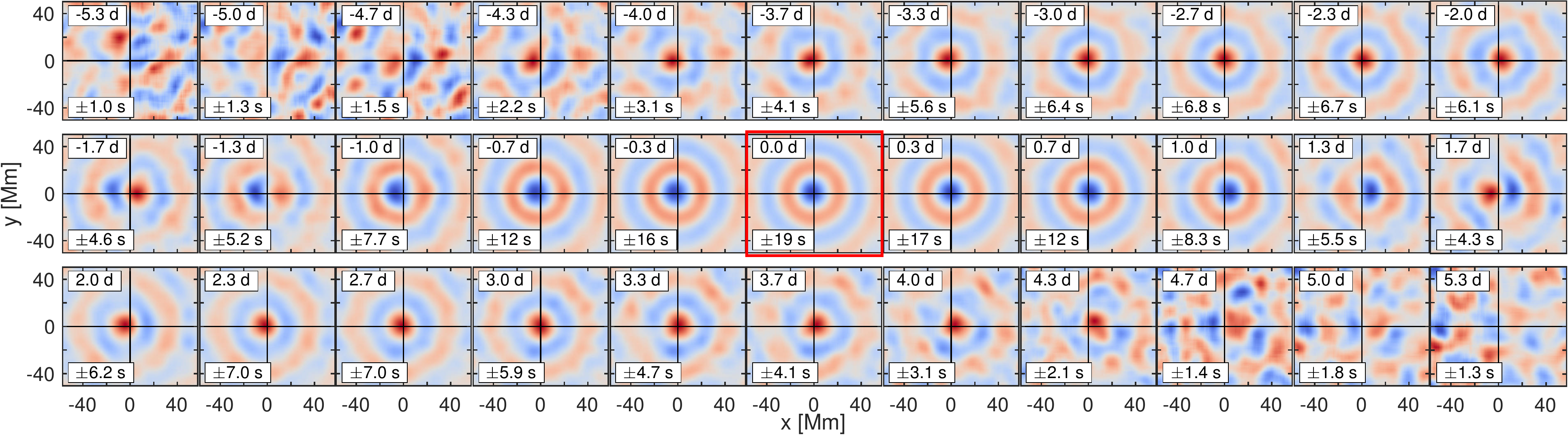} \\
\includegraphics[width=\hsize]{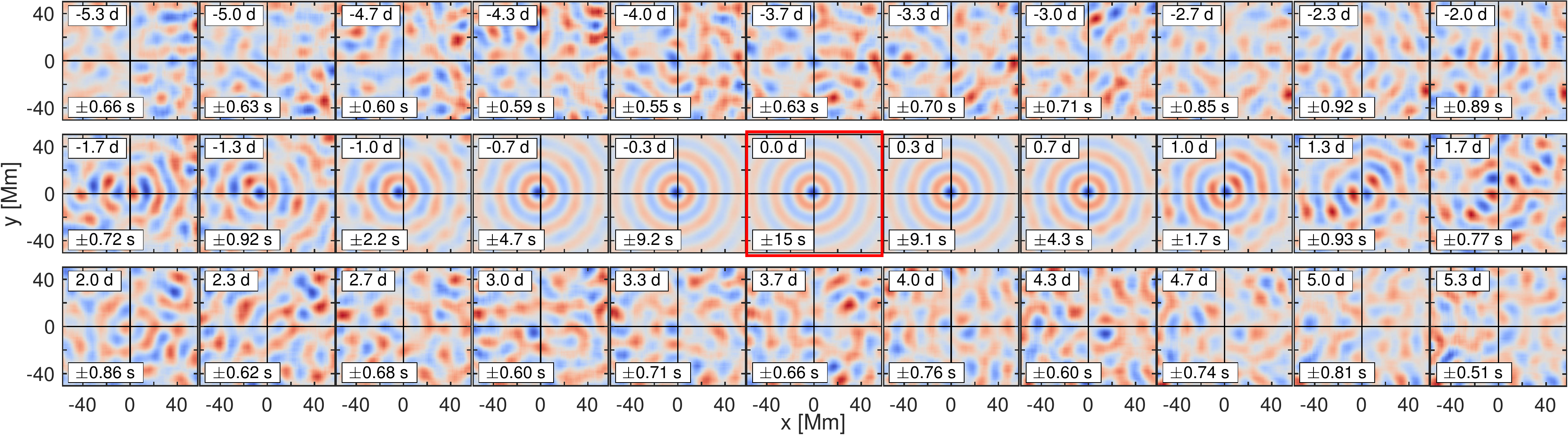}
\caption{Evolution of f-mode-filtered inward$-$outward travel times for the average convective inflow (cf.~Fig.~\ref{fig_tauoi}, bottom panels) at three different scales. \textit{From top to bottom:} $kR_\odot = 52$ (${\sim}$80~Mm), $kR_\odot = 124$ (${\sim}$35~Mm), $kR_\odot = 222$ (${\sim}$20~Mm). The spatial band-pass filters are Gaussian in $kR_\odot$ with $\sigma \approx 12$.}
\label{fig_tauoi_kRscales_inflow}
    \end{figure*}

%______________________________________________ LCT div_h evolution for avg. SG outflow (kR=52, 124, 222)
   \begin{figure*}[h]
\centering
\includegraphics[width=\hsize]{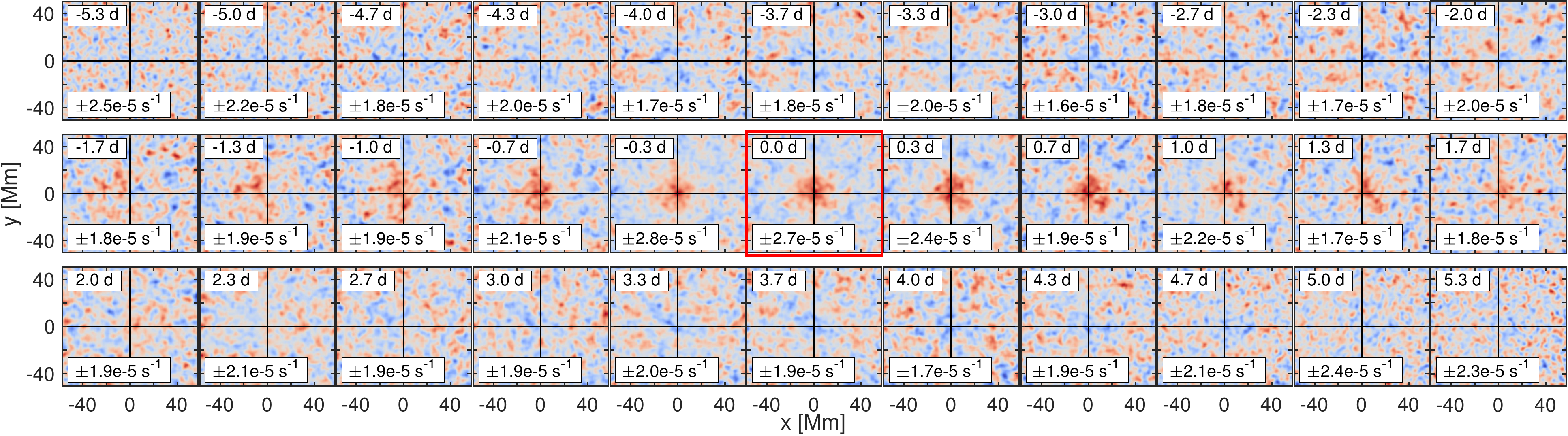} \\
\includegraphics[width=\hsize]{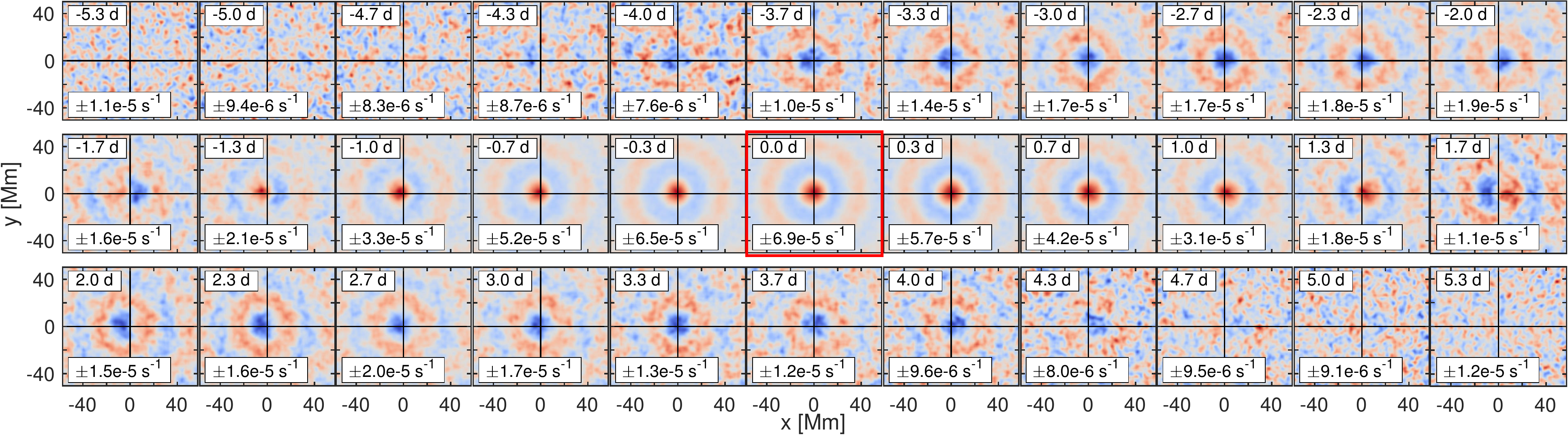} \\
\includegraphics[width=\hsize]{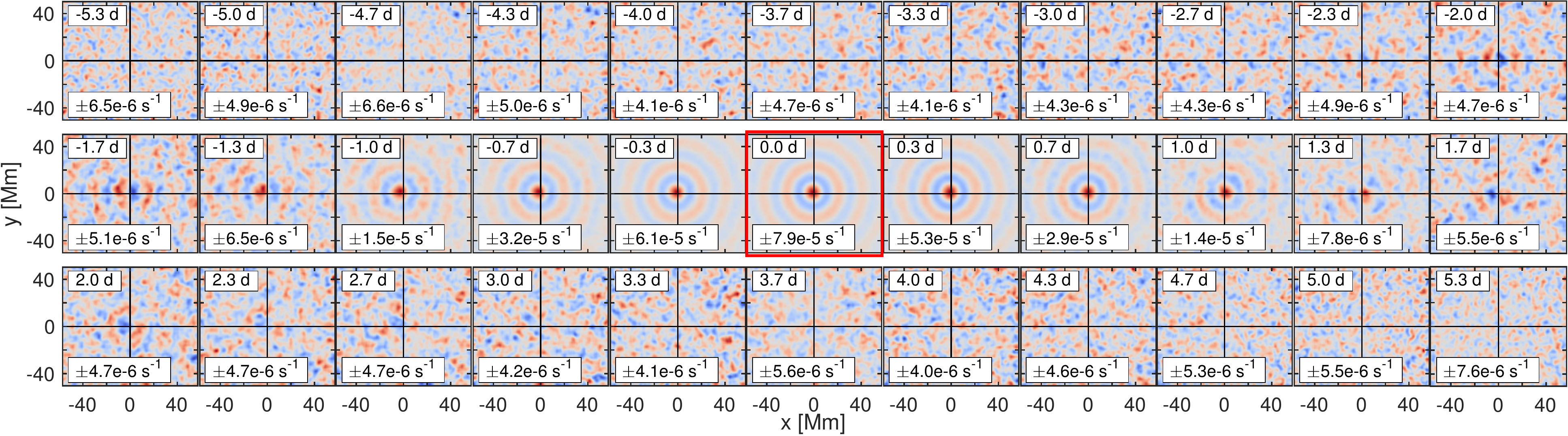}
\caption{Evolution of LCT divergence for the average supergranular outflow (cf.~Fig.~\ref{fig_LCTdiv}) at three different scales. \textit{From top to bottom:} $kR_\odot = 52$ (${\sim}$80~Mm), $kR_\odot = 124$ (${\sim}$35~Mm), $kR_\odot = 222$ (${\sim}$20~Mm).}
\label{fig_lctdiv_kRscales_outflow}
    \end{figure*}

%______________________________________________ LCT div_h evolution for avg. SG inflow (kR=52, 124, 222)
   \begin{figure*}[h]
\centering
\includegraphics[width=\hsize]{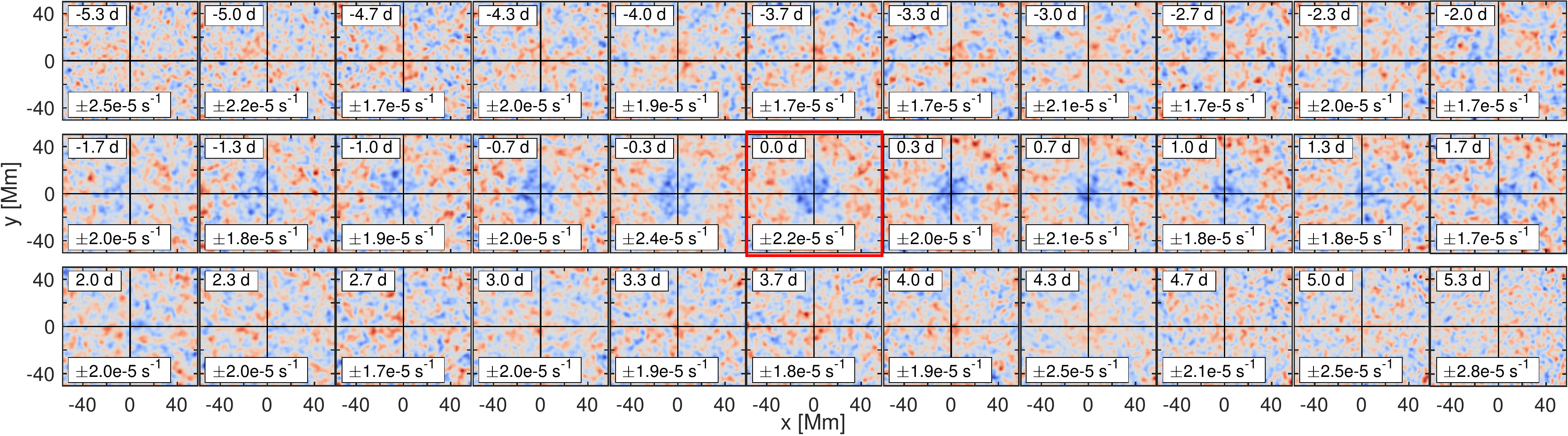} \\
\includegraphics[width=\hsize]{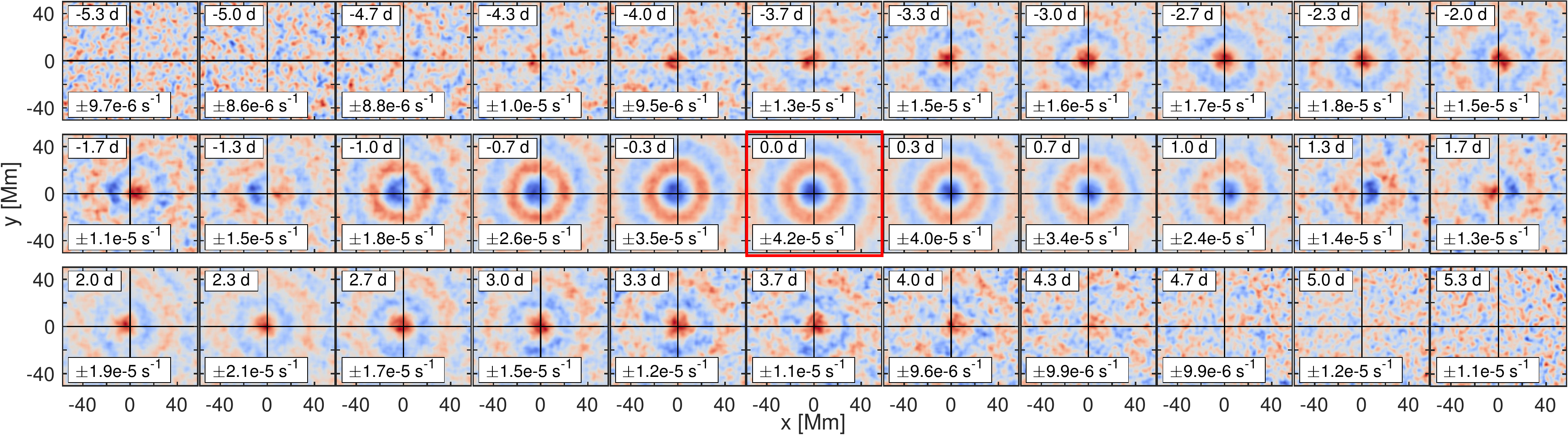} \\
\includegraphics[width=\hsize]{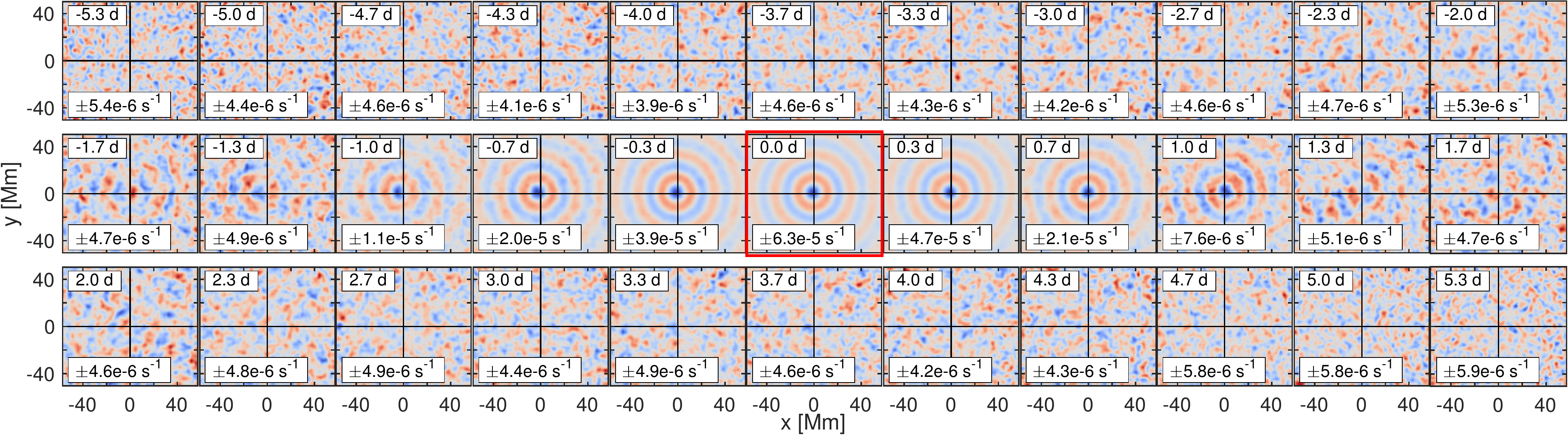}
\caption{Evolution of LCT divergence for the average supergranular inflow at three different scales. \textit{From top to bottom:} $kR_\odot = 52$ (${\sim}$80~Mm), $kR_\odot = 124$ (${\sim}$35~Mm), $kR_\odot = 222$ (${\sim}$20~Mm).}
\label{fig_lctdiv_kRscales_inflow}
    \end{figure*}

%______________________________________________ tau^oi evolution: y-band average inflow (space-time image)
    \begin{figure*}[h]
\centering
\includegraphics[width=\textwidth]{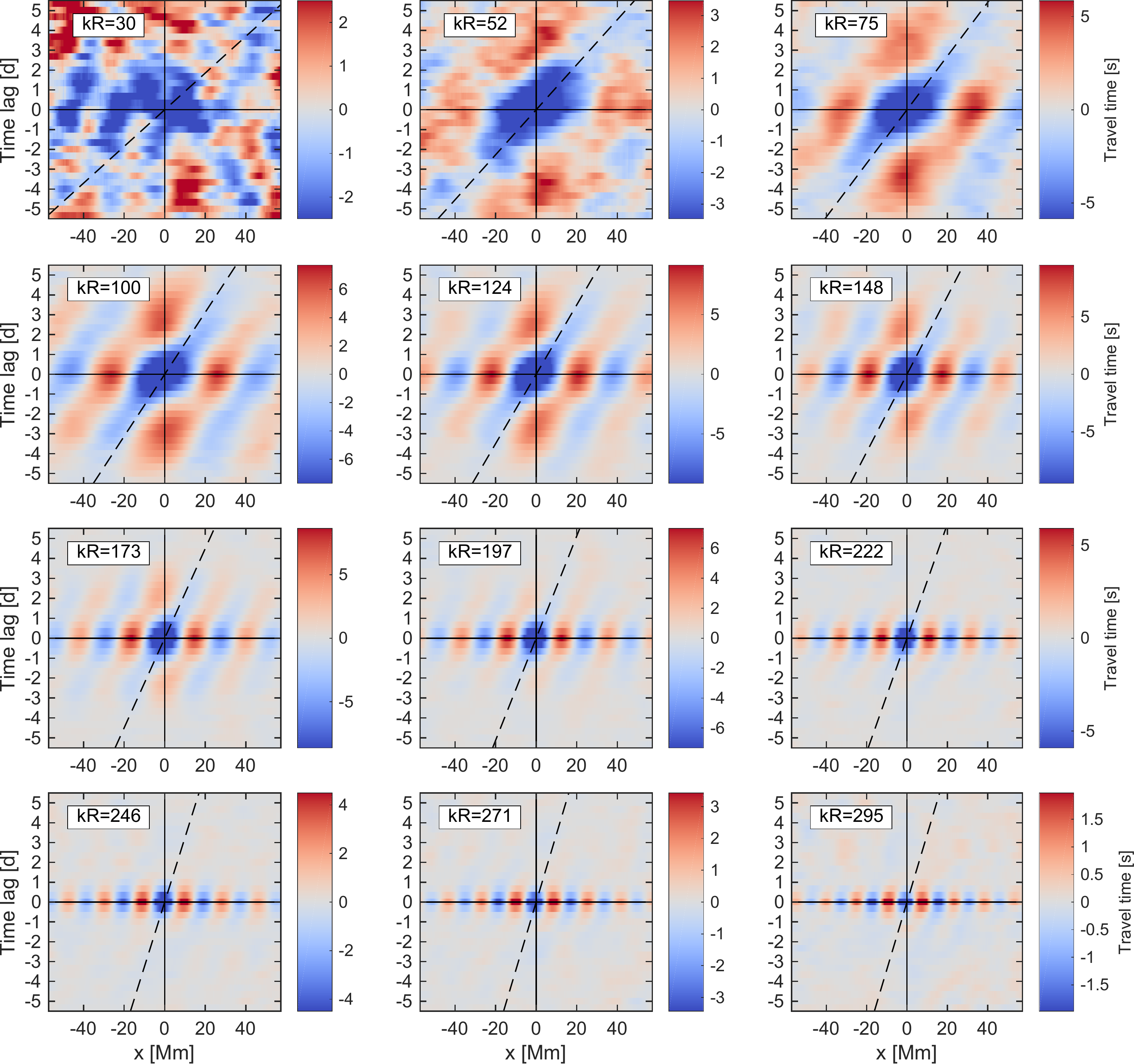}
\caption{As Fig.~\ref{fig_evolution-yavg}, but for the average inflow.}
\label{fig_evolution-yavg-in}
    \end{figure*}

%______________________________________________ LCT evolution: y-band average outflow (space-time image)
   \begin{figure*}[h]
\centering
\includegraphics[width=\textwidth]{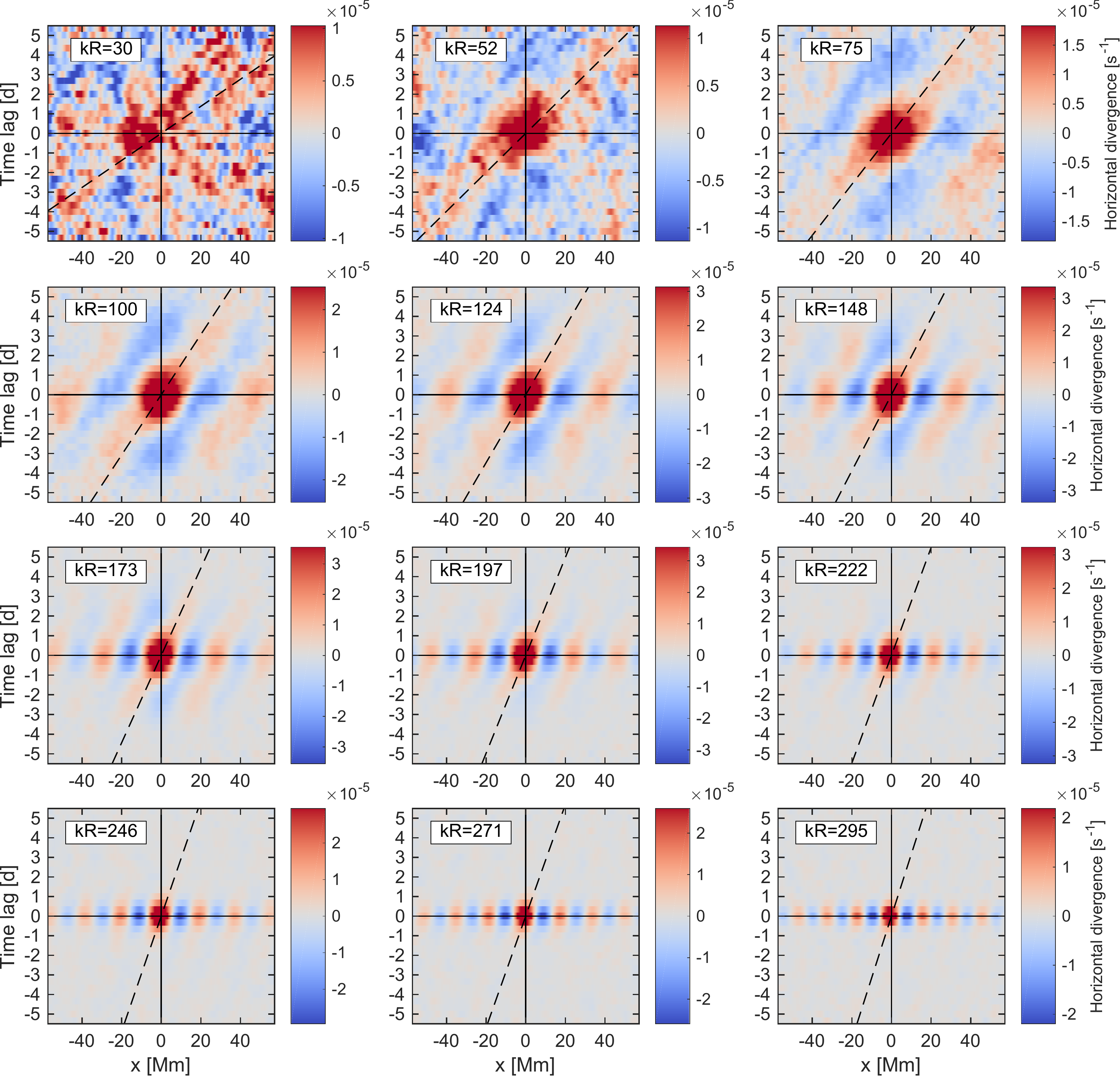}
\caption{Evolution for different scales: space-time image. Average over y band (10 Mm) for the average outflow using LCT. The dashed lines give the phase speed from the dispersion relation as obtained from the power spectrum.}
\label{fig_evolution-yavg-lct}
    \end{figure*}

%______________________________________________ LCT evolution: y-band average inflow (space-time image)
   \begin{figure*}[h]
\centering
\includegraphics[width=\textwidth]{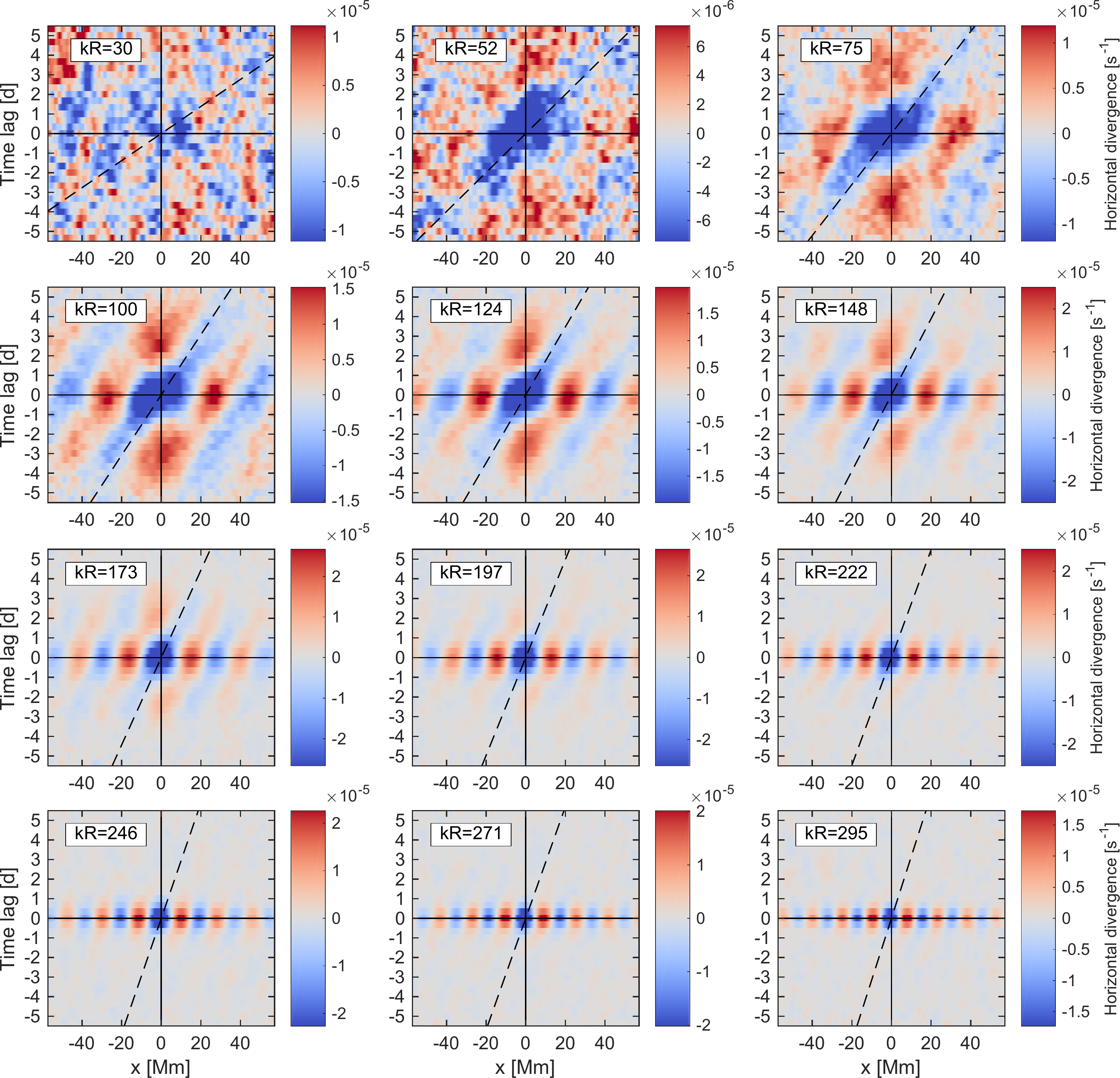}
\caption{As Fig.~\ref{fig_evolution-yavg-lct}, but for the average inflow.}
\label{fig_evolution-yavg-lct-in}
    \end{figure*}

\end{document}